\documentclass[
 reprint,
 amsmath,amssymb,
 aps,
 prd,
]{revtex4-2}

\usepackage{dcolumn}
\usepackage{indentfirst}
\usepackage{amsmath}
\usepackage{bm}
\usepackage{slashed}
\usepackage{graphicx}
\usepackage{hyperref}
\usepackage{multirow}

\begin{document}

\preprint{APS/123-QED}

\title{Cosmological Constraints on Thermal Friction of Axion Dark Matter}

\author{Gang Liu}
 \email{liugang\_dlut@mail.dlut.edu.cn}

\author{Yuhao Mu}%
\author{Zhihuan Zhou}%

\author{Lixin Xu}
 \email{lxxu@dlut.edu.cn}
\affiliation{%
 Institute of Theoretical Physics\\
 School of Physics\\
 Dalian University of Technology\\
 Dalian 116024, People's Republic of China
}

\date{\today}

\begin{abstract}
In this paper, we investigate the process in which axion dark matter undergoes thermal 
friction, resulting in energy injection into dark radiation, with the aim of mitigating
the Hubble tension and large-scale structure tension. In the early universe, this 
scenario led to a rapid increase in the energy density of dark radiation; 
in the late universe, the evolution of axion dark matter is similar to that of 
cold dark matter, with this scenario resembling decaying dark matter and serving to 
ease the large-scale structure tension. We employ cosmological observational data, 
including cosmic microwave background (CMB), baryon acoustic oscillation (BAO), 
supernova data (SNIa), $H_0$ measurement from SH0ES, and $S_8$ from the Dark Energy 
Survey Year-3 (DES), to study and analyze this model. Our results indicate that the 
thermal friction model offers partial alleviation of the large-scale structure tension, 
while its contribution on alleviating Hubble tension can be ignored. 
The new model yields the value of $S_8$ is $0.795\pm 0.011$ at a 68\% confidence level, 
while the $\Lambda$CDM model yields a result of $0.8023\pm 0.0085$. 
In addition, the new model exhibits a lower $\chi^2_\mathrm{tot}$ value, with 
a difference of -2.60 compared to the $\Lambda$CDM model. 
Additionally, we incorporate Lyman-$\alpha$ data to re-constrain the new model and find 
a slight improvement in the results, with the values of $H_0$ and $S_8$ being 
$68.76^{+0.39}_{-0.35}$ km/s/Mpc and $0.791\pm 0.011$ at a 68\% confidence level, respectively.
\end{abstract}

\maketitle


\section{Introduction}
In recent years, an increasing number of observational facts have shown inconsistencies 
with the predicted results of the $\Lambda$CDM model, notably the Hubble and large-scale 
structure tensions, which may indicate the presence of new physics beyond the the 
standard cosmological model. 
The Hubble tension refers to the discrepancy between the predicted value of the Hubble 
constant, $H_0$, from the $\Lambda$CDM model and the observed value \cite{Verde_2019, freedman2017}. 
The \textit{Planck} 2018 CMB data yielded a result of $67.37\pm0.54$ km/s/Mpc \cite{planck2020}, 
whereas the SH0ES measurement obtained via the distance ladder method at low redshifts 
resulted in a value of $73.04\pm1.04$ km/s/Mpc \cite{Riess2021ACM}, with a statistical 
error of 4.8$\sigma$. 

The large-scale structure displays a moderate level of tension with CMB, typically quantified by 
$S_8\equiv\sigma_8\sqrt{(\Omega_m/0.3)}$, where $\sigma_8$ represents the amplitude 
of density perturbations, and $\Omega_m$ denotes the matter density fraction. 
The \textit{Planck} 2018 best-fit $\Lambda$CDM 
model yields a value of $0.834\pm0.016$ for $S_8$ \cite{planck2020}; however, 
observational results from large-scale structure surveys such as the Dark Energy 
Survey Year-3 (DES) suggest a lower value of $0.776\pm0.017$ \cite{PRD.105.023520}.

Various models have been proposed to address these inconsistencies. Common ones include 
modifying dark energy, including dynamic dark energy \cite{Li_2019, Zhouzh, Guo_2019}, 
vacuum phase transition \cite{PRD.97.043528, PRD.99.083509, PRD.101.123521}, early dark 
energy \cite{PRL.122.221301, PRD.101.063523, PRD.102.043507}, and interacting dark energy 
\cite{sty2789, Di_Valentino_2020, sty2780, liu2023alleviating, liu2023kinetically, liu2023mitigating}; 
modifying dark matter, including partially acoustic dark matter \cite{PRD.96.103501}, 
decaying dark matter \cite{PRD.99.121302, Buch_2017, PRD.98.023543, PRD.103.043014, 
Pandey_2020, Alvi_2022, mccarthy2023converting}, interacting dark matter \cite{PRD.105.103509}, 
and axion dark matter \cite{D'Eramo_2018}. In addition, modification of the gravitational 
theory is also considered \cite{El-Zant_2019, PRD.99.103526, Renk_2017, PRD.102.023529}. 

In this paper, we focus on the interaction between dark matter and dark radiation. 
We investigate this interaction by substituting cold dark matter with 
axion dark matter, and examine the problem from the perspective of scalar field theory. 
We consider the impact of additional thermal friction on axion dark matter, which 
injects energy into dark radiation and alters the evolution equations in the original 
model of axion dark matter.

Axion dark matter is a potential candidate for dark matter, which can form 
Bose-Einstein condensates on galactic scales, altering the dynamics of dark matter 
on small scales while maintaining the success of cold dark matter on large scales 
\cite{PRD.106.123501}. In this paper, we investigate the model of ultra-light axion 
dark matter, with a mass $m_{\chi}$ approximately at $10^{-22}$ eV. This particular dark matter 
model has been extensively studied in previous works \cite{2005.03254, Ure_a_L_pez_2016, 
PRD.96.061301}.

\begin{figure}
    \includegraphics[width=\columnwidth]{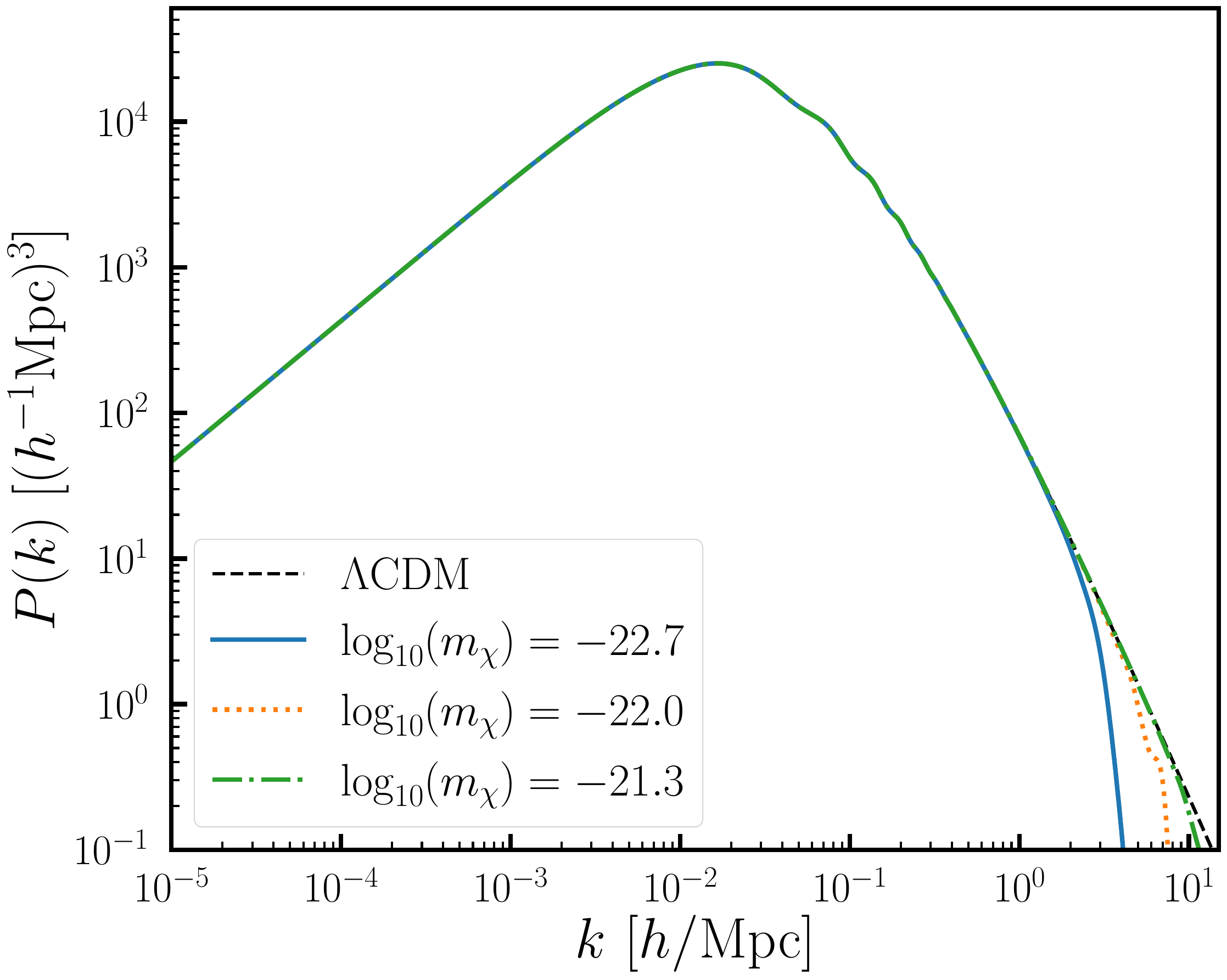}
    \caption{\label{fig1}The impact of axion mass on the matter power spectrum. On large 
    scales, axion dark matter behaves similar to cold dark matter, while on small scales, 
    it exhibits decreased clustering due to condensation. The smaller the axion 
    mass, the more pronounced this effect becomes.}
\end{figure}

Figure~\ref{fig1} presents the influence of axion mass on the matter power spectrum. 
We keep other cosmological parameters fixed and consistent with the $\Lambda$CDM model.
On large scales, the evolution of axion dark matter is similar to that of cold dark 
matter. However, on small scales, the condensation effect of axion dark matter inhibits 
structure growth, resulting in a smaller matter power spectrum than that predicted by 
the $\Lambda$CDM model. Furthermore, this effect becomes more apparent as the axion mass 
decreases.

In the early universe, the axion field was frozen due to Hubble 
friction. As the universe cooled and the Hubble parameter decreased, the axion mass 
became comparable to the Hubble friction, causing the axion to start rolling and 
subsequently oscillating near the minimum of its potential. After the axion potential 
and kinetic energies reach equilibrium, its energy density evolves like 
that of matter \cite{liu2023alleviating, liu2023kinetically}. 

We extend the axion dark matter model by proposing 
the ``dissipative axion dark matter'' (DADM) model, which takes into account the coupling 
between axion and non-abelian gauge group. This results in scalar field experiencing 
thermal friction beyond Hubble friction, transferring energy to the dark radiation 
composed of dark gauge bosons.

The thermal friction effect of axion has been studied in the context of warm inflation 
\cite{PRL.75.3218, PRL.74.1912, Berera_2009, PRL.117.151301, Berghaus_2020}, late-time 
dark energy \cite{PRD.100.015048, PRD.104.083520}, and early-time dark energy 
\cite{PRD.101.083537, berghaus2022thermal}. In this study, we apply it to the realm of 
axion dark matter.

We constrained the parameters of the DADM model using a combination of various 
cosmological datasets, including cosmic microwave background (CMB), baryon acoustic 
oscillation (BAO), supernova data (SNIa), $H_0$ measurement from SH0ES, and $S_8$ from 
the Dark Energy Survey Year-3 (DES). 

The results indicate that the new model yields the value of $H_0$ that is close to the 
result obtained from the $\Lambda$CDM model, while the $S_8$ in the former is smaller 
than in the latter. The DADM model 
yields the value of $S_8$ is $0.795\pm 0.011$ at a 68\% confidence level, while the 
$\Lambda$CDM model produces a result of $0.8023\pm 0.0085$. 
This suggests that the thermal friction model can alleviate the 
$S_8$ tension. 
Additionally, the DADM model demonstrates a lower $\chi^2_\mathrm{tot}$ value, 
differing by -2.60 compared to the $\Lambda$CDM model.

In the analysis of the constraint results, we discovered that the utilization of only 
the aforementioned five datasets is insufficient to constrain the mass of axion dark 
matter. Therefore, we fixed the mass of the axion dark matter to be $10^{-22}$ eV and 
incorporated Lyman-$\alpha$ data to re-constrain the new model. The results show some 
improvements, with values of $H_0$ and $S_8$ being 
$68.76^{+0.39}_{-0.35}$ km/s/Mpc and $0.791\pm 0.011$ at a 68\% confidence level, respectively

The present paper is structured as follows. In Sec.~\ref{sec2}, we present the 
thermal friction model, deduce the evolution equations for both the background and 
perturbation, and provide the corresponding initial conditions. Section \ref{sec3} 
presents the numerical analysis, including the effects of the model on the background 
and perturbation levels. In Sec.~\ref{sec4}, we describe the data constrained by 
Markov Chain Monte Carlo (MCMC) analyses and provide the results. Finally, in 
Sec.~\ref{sec5}, we discuss the findings and offer a conclusion.

\section{\label{sec2}Thermal Friction Model}
We extend the Standard Model by introducing a dark non-Abelian gauge group and axion 
field $\chi$. The dark gauge boson couple to the axion via the operator, 
\begin{equation}
    \mathcal{L}_\mathrm{int}=\frac{\alpha}{16\pi}\frac{\chi}{f}\tilde{F}^{\mu\nu}_aF^a_{\mu\nu}, 
\end{equation}
where $F^a_{\mu\nu}$($\tilde{F}^{\mu\nu}_{a}=\epsilon^{\mu\nu\alpha\beta}F^{a}_{\alpha\beta}$) 
represents the field strength of the dark gauge boson, $\alpha$ is the fine 
structure constant, and $f$ denotes the strength of the interaction between axion and 
the dark gauge field. We consider a simple quadratic potential for the axion, 
\begin{equation}
    V(\chi)=\frac{1}{2}m_{\chi}^2\chi^2, 
\end{equation}
while various forms of axion potential have been studied in previous works 
\cite{BEYER2014418,AMENDOLA2006192}, which can be effectively represented by this 
quadratic potential \cite{Ure_a_L_pez_2016}. We assume that these two dark sector 
components interact with other components only through gravity. The conservation of 
energy-momentum tensor holds for two components, 
\begin{equation}
    \nabla_{\mu}T^{\mu\nu}_{\chi}=-\nabla_{\mu}T^{\mu\nu}_\mathrm{dr}. 
\end{equation}
Using the results from 
\cite{Bastero-Gil_2014, berghaus2022thermal}, we have, 
\begin{equation}
    \nabla_{\mu}T^{\mu\nu}_\mathrm{dr}=g^{\nu\alpha}(-\Upsilon v^{\mu}_\mathrm{dr}\partial_{\mu}\chi\partial_{\alpha}\chi), \label{emt}
\end{equation}
where $v^{\mu}_\mathrm{dr}$ represents the four-velocity of the dark radiation. The 
thermal friction coefficient $\Upsilon$ is expected to be correlated with the 
temperature of dark radiation \cite{PRD.43.2027, Moore_2011, Laine_2016}. However, in 
this paper, we treat $\Upsilon$ as a constant for simplicity and restrict our 
analysis to the case of weak thermal friction. Investigation of more complex 
scenarios is left to future work.

\subsection{Background Equations}
The background evolution equations for the axion dark matter and dark radiation can be derived 
using Eq.(\ref{emt}), 
\begin{subequations}
    \begin{align}
        &\ddot{\chi}+(3H+\Upsilon)\dot{\chi}+m_{\chi}^2\chi=0, \\
        &\dot{\rho}_\mathrm{dr}+4H\rho_\mathrm{dr}=\Upsilon\dot{\chi}^2 \label{rhodr}, 
    \end{align} 
\end{subequations}
where the dot denotes the derivative with respect to cosmic time, and $H$ is the
Hubble parameter. The motion equation of a scalar field includes thermal friction, 
which results in the transfer of energy from axion to the dark radiation.

The energy density and pressure of the axion dark matter can be expressed as follows, 
\begin{subequations}
    \begin{align}
        &\rho_{\chi}=\frac{1}{2}\dot{\chi}^2+\frac{1}{2}m_{\chi}^2\chi^2,\\
        &p_{\chi}=\frac{1}{2}\dot{\chi}^2-\frac{1}{2}m_{\chi}^2\chi^2.
    \end{align} 
\end{subequations}
We introduce new variables to calculate the equation of motion for the scalar field 
\cite{Ure_a_L_pez_2016, PRD.57.4686, PRD.96.061301},
\begin{subequations}
    \begin{align}
        \sqrt{\Omega_{\chi}}\sin{\frac{\theta}{2}}&=\frac{\dot{\chi}}{\sqrt{6}M_{pl}H},\\
        \sqrt{\Omega_{\chi}}\cos{\frac{\theta}{2}}&=-\frac{m_{\chi}\chi}{\sqrt{6}M_{pl}H},\\
        y_1&=\frac{2m_{\chi}}{H},
    \end{align} 
\end{subequations}
where $\Omega_{\chi}=\frac{\rho_{\chi}}{3M_{pl}^2H^2}$ is the density parameter 
of the dark matter and $M_{pl}=2.435\times10^{27}$ eV denotes the reduced Planck mass. 
The evolution equations for the new variable are, 
\begin{subequations}
    \begin{align}
        &\frac{\dot{\Omega}_{\chi}}{\Omega_{\chi}}=3H(w_t+\cos\theta)-\Upsilon(1-\cos\theta),\\
        &\dot{\theta}=Hy_1-(3H+\Upsilon)\sin\theta,\\
        &\dot{y}_1=\frac{3}{2}Hy_1(1+w_t),
    \end{align} 
\end{subequations}
where $w_t$ represents the total equation of state, 
which is the ratio of total pressure to total energy density.
We can find that the equation of state of axion dark matter is, 
\begin{equation}
    w_{\chi}=\frac{p_{\chi}}{\rho_{\chi}}=-\cos\theta.
\end{equation}
The continuity equations for axion dark matter and dark radiation can be derived as 
follows, 
\begin{subequations}
    \begin{align}
        &\dot{\rho}_{\chi}+3H\rho_{\chi}=-\Upsilon\rho_{\chi}(1+w_{\chi}), \\
        &\dot{\rho}_\mathrm{dr}+4H\rho_\mathrm{dr}=\Upsilon\rho_{\chi}(1+w_{\chi}). 
    \end{align} 
\end{subequations}
After the cessation of axion oscillation (which occurs during the radiation domination 
epoch), its equation of state $w_{\chi}=0$, and the continuity equation takes on the 
same form as that of decaying dark matter \cite{PRD.98.023543, PRD.103.043014, 
Pandey_2020, Alvi_2022, mccarthy2023converting, 
Audren_2014, Lesgourgues_2016}. 
However, our result is obtained from the 
scalar field undergoing thermal friction. 

\begin{figure}
    \includegraphics[width=\columnwidth]{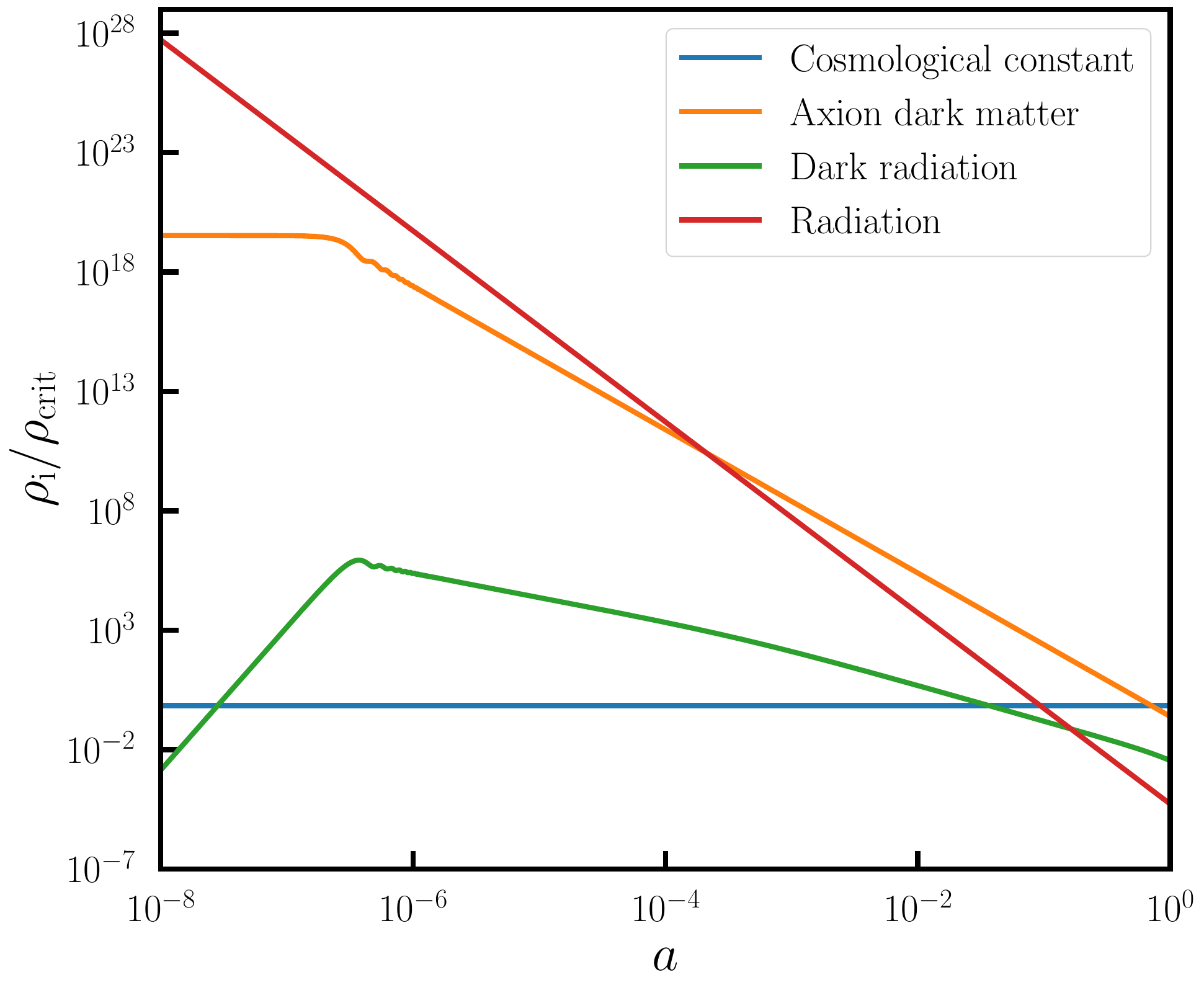}
    \caption{\label{fig2}The energy density of various components of the universe varies 
    with the scale factor, including ordinary radiation (red line), axion dark matter 
    (orange line), dark radiation (green line), and cosmological constant (blue line). 
    The dark radiation generated by axion thermal friction only has an impact in the 
    late time of universe. }
\end{figure}

Figure~\ref{fig2} 
depicts the variation of energy density for different components 
as a function of scale factor. The orange curve represents axion dark matter, which 
behaves similar to a cosmological constant (blue line) in the early universe and evolves 
like matter after undergoing oscillations. Dark radiation (green line) generated by axion 
thermal friction is negligible in the early universe. However, due to a continuous 
source, the dilution rate of dark radiation is slower than that of ordinary radiation 
(red line) and matter. 

\subsection{Perturbution Equations}
We calculated the perturbation equations for axion dark matter and dark radiation with 
the synchronous gauge, where the line element is defined as, 
\begin{equation}
    ds^2=-dt^2+a^2(t)(\delta_{ij}+h_{ij})dx^idx^j.
\end{equation}
The perturbed Klein-Gordon equation for a scalar field in Fourier modes is expressed 
as follows:
\begin{equation}
    \ddot{\delta\chi}+(3H+\Upsilon)\dot{\delta\chi}+(\frac{k^2}{a^2}+m_{\chi}^2)\delta\chi
    +\frac{1}{2}\dot{h}\dot{\chi}=0,
\end{equation}
where $h$ representes the trace of scalar metric perturbations.
In accordance with \cite{PRD.58.023503,Hu_1998}, the density perturbation, pressure 
perturbation, and velocity divergence of axion dark matter can be expressed as,
\begin{subequations}
    \begin{align}
        &\delta\rho_{\chi}=\dot{\chi}\dot{\delta\chi}+m_{\chi}^2\chi\delta\chi,\\
        &\delta p_{\chi}=\dot{\chi}\dot{\delta\chi}-m_{\chi}^2\chi\delta\chi,\\
        &(\rho_{\chi}+p_{\chi})\Theta_{\chi}=\frac{k^2}{a}\dot{\chi}\delta\chi.
    \end{align} 
\end{subequations}

We redefine new variables to calculate the perturbation equation of scalar field \cite{PRD.96.061301},
\begin{subequations}
    \begin{align}
        &\sqrt{\Omega_{\chi}}(\delta_0\sin{\frac{\theta}{2}}+\delta_1\cos{\frac{\theta}{2}})=\sqrt{\frac{2}{3}}\frac{\dot{\delta\chi}}{M_{pl}H},\\
        &\sqrt{\Omega_{\chi}}(\delta_1\sin{\frac{\theta}{2}}-\delta_0\cos{\frac{\theta}{2}})=\sqrt{\frac{2}{3}}\frac{m_{\chi}\delta\chi}{M_{pl}H}.
    \end{align} 
\end{subequations}

The evolution equations of the new variable can be derived as follows, 
\begin{subequations}
    \begin{align}
        \dot{\delta}_{0}&=\delta_0H\omega\sin\theta-\delta_1[(3H+\Upsilon)\sin\theta\\
        \notag 
        &+H\omega(1-\cos\theta)]-\frac{\dot{h}}{2}(1-\cos\theta),\\
        \dot{\delta}_{1}&=\delta_0H\omega(1+\cos\theta)-\delta_1((3H+\Upsilon)\cos\theta\\
        \notag
        &+H\omega\sin\theta)-\frac{\dot{h}}{2}\sin\theta,
    \end{align} 
\end{subequations}
where 
\begin{equation}
    \omega=\frac{k^2}{2a^2m_{\chi}H}=\frac{k^2}{a^2H^2y_1}.
\end{equation}
Employing Eq.(\ref{emt}), we derive the perturbed energy density and velocity evolution 
equations for dark radiation, 
\begin{small}
    \begin{subequations}
        \begin{align}
            &\dot{\delta}_\mathrm{dr}=-\frac{2}{3}\dot{h}-\frac{4}{3a}\Theta_\mathrm{dr}+\Upsilon\frac{\rho_{\chi}}{\rho_\mathrm{dr}} \left[
            (\delta_0-\delta_\mathrm{dr})(1-\cos\theta)+\delta_1\sin\theta\right],\\
            &\dot{\Theta}_\mathrm{dr}=\frac{1}{4a}k^2\delta_\mathrm{dr}+\Upsilon\frac{\rho_{\chi}}{\rho_\mathrm{dr}}(1-\cos\theta)(\frac{3}{4}\Theta_{\chi}-\Theta_\mathrm{dr}).
        \end{align} 
    \end{subequations}
\end{small}

It is worth noting that the dark radiation 
generated by thermal friction experiences sufficient self-interaction to maintain a 
thermal environment and suppress shear perturbations \cite{berghaus2022thermal}. In 
contrast, in most decaying 
dark matter models, the dark radiation exhibits shear perturbations that 
cannot be neglected due to the free-streaming \cite{mccarthy2023converting, 
Lesgourgues_2016, Poulin_2016, PRD.106.023516}.

\subsection{Initial Conditions}
In the very early universe, the Hubble friction dominates greatly over both the axion 
mass and thermal friction. When studying the initial conditions of the axion dark matter, 
the influence of $\Upsilon$ can be neglected, and the scalar field equation simplifies 
to the case without thermal friction. Therefore, we can directly employ the initial 
conditions for the new variable of the scalar field presented in \cite{Ure_a_L_pez_2016, PRD.96.061301}.

We assume that dark radiation is exclusively generated through thermal friction of axion 
dark matter. Therefore, the initial value of $\rho_\mathrm{dr}$ is set to zero.
For the perturbation equations, we use adiabatic initial conditions,
\begin{equation}
    \delta_\mathrm{dr}=\frac{3}{4}\delta_{\gamma},\quad 
    \Theta_\mathrm{dr}=\Theta_{\gamma}, 
\end{equation}
where the subscript ``$\gamma$'' represents the photon.

\section{\label{sec3}Numerical Results}
Based on the description in the last section, we modified the publicly available 
Boltzmann code $\mathtt{CLASS}$ \cite{1104.2932,Blas_2011}. We now present 
the obtained numerical results and discuss the impact of the dissipative axion dark 
matter (DADM) model on observations. 
We employ the parameter values of the $\Lambda$CDM model constrained by \textit{Planck} 
2018 data, 
\begin{align}
    &\omega_\mathrm{b}=0.02238, \qquad \omega_\mathrm{cdm}=0.1201,\\
    \notag
    &H_0=67.81, \qquad \ln(10^{10}A_\mathrm{s})=3.0448,\\
    \notag
    &n_\mathrm{s}=0.9661, \qquad \tau_\mathrm{reio}=0.0543.
\end{align}
For the DADM model, we set the axion mass to be $10^{-22}$ eV and keep all other 
parameters the same as the $\Lambda$CDM model, except for the sum of the dark matter 
and dark radiation energy density, $\omega_\mathrm{dm+dr}$. 

In order to facilitate model comparison, we fix the axion dark matter energy 
density fraction $\Omega_\mathrm{dm}(z)$ at the epoch of matter-radiation equality is equal 
to the cold dark matter energy density fraction $\Omega_\mathrm{cdm}(z)$ in the 
$\Lambda$CDM model. This can be achieved by adjusting the value of $\omega_\mathrm{dm+dr}$ 
after setting the thermal friction coefficient $\Upsilon$.

We compare the results obtained with thermal friction coefficients $\Upsilon$ of 0, 1.5, 
and 2.5 to those of the $\Lambda$CDM model. The four cases are represented by solid blue, 
dotted orange, dash-dotted green, and dashed black lines, respectively. The units of 
$\Upsilon$ are consistent with $H_0$, i.e., km/s/Mpc.

\begin{figure}
    \includegraphics[width=\columnwidth]{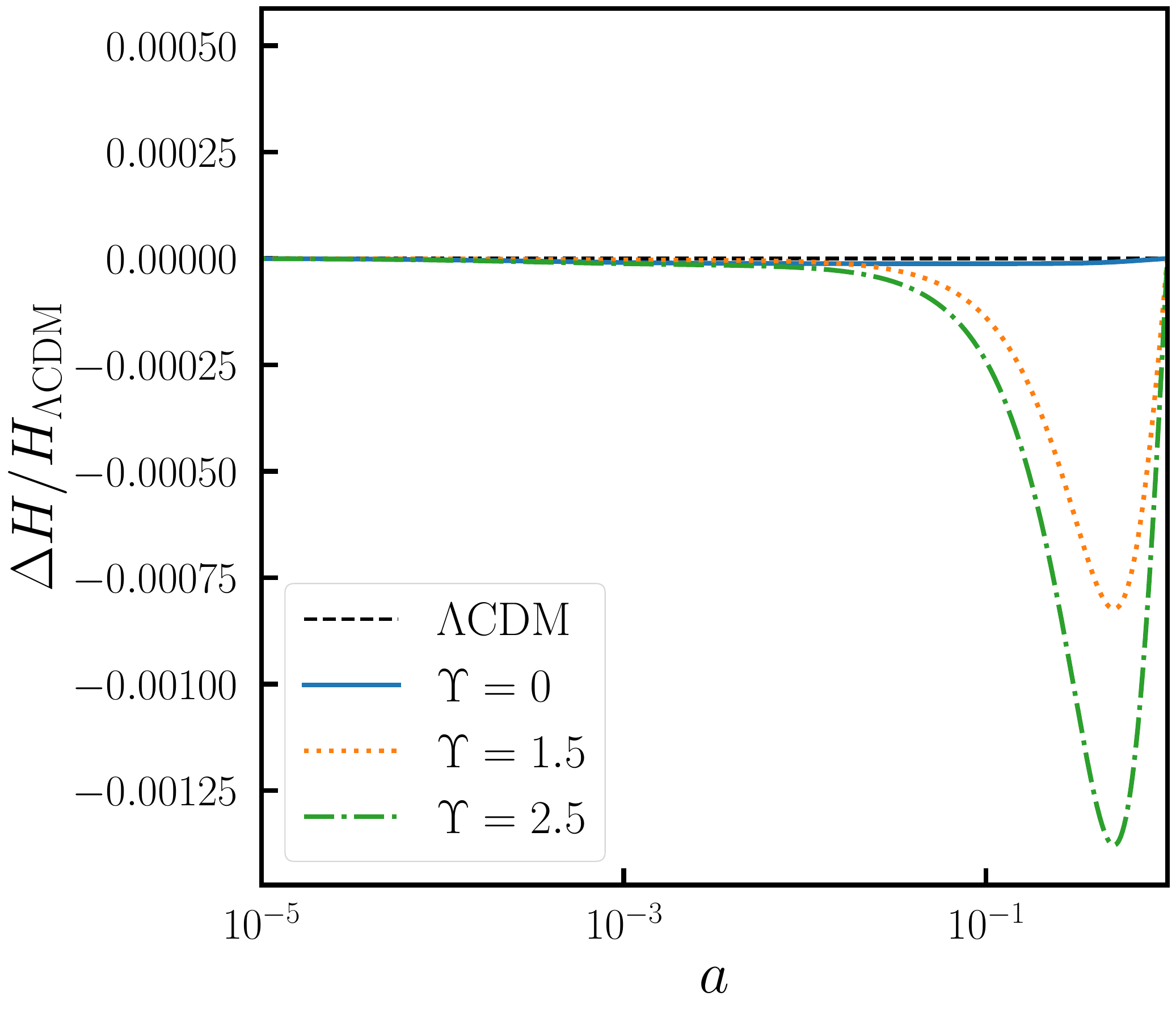}
    \caption{\label{fig3}The evolution of the Hubble parameter relative to the 
    $\Lambda$CDM model as a function of the scale factor is depicted for the DADM model 
    under different thermal friction coefficients. The interaction between dark matter 
    and dark radiation causes the expansion rate during the matter-dominated era in the 
    new model to be smaller than that of the $\Lambda$CDM model. }
\end{figure}

In Fig.~\ref{fig3}, the evolution of the Hubble parameter relative to the $\Lambda$CDM 
model as a function of the scale factor for different thermal friction coefficients are 
presented. In the DADM model, due to the continuous energy transfer from dark matter to 
dark radiation, the expansion rate during the matter-dominated era is smaller compared 
to the $\Lambda$CDM model. This trend persists until the dark energy domination era 
takes over. Moreover, the larger the thermal friction coefficient, the greater the 
disparity between the two models.

\begin{figure}
    \includegraphics[width=\columnwidth]{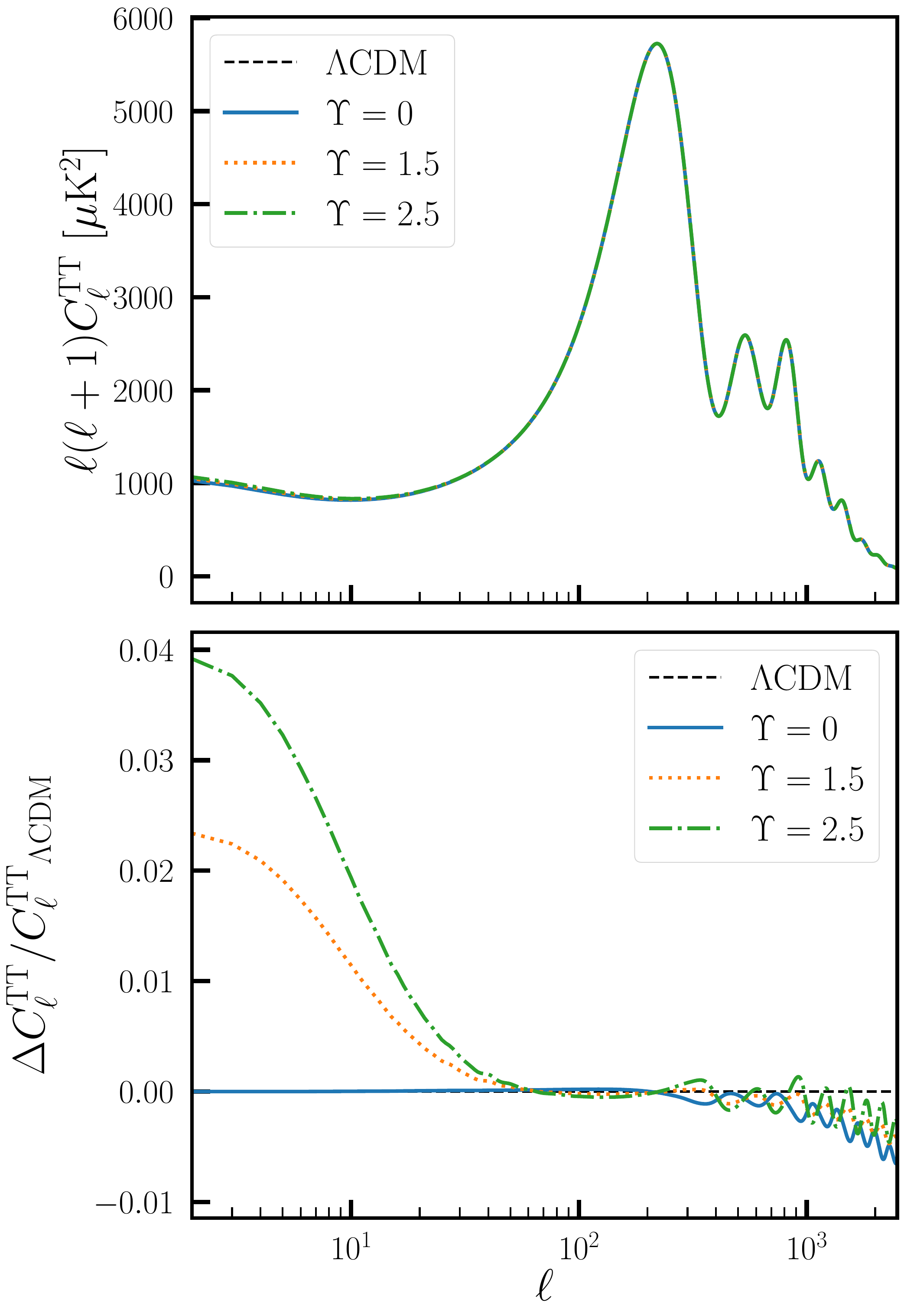}
    \caption{\label{fig4}The impact of various thermal friction coefficients on the 
    CMB temperature power spectrum. 
    The DADM model exhibits a notable impact on the CMB power spectra at low $\ell$, with 
    the spectrum significantly enhanced compared to the $\Lambda$CDM 
    model. This variation primarily stems from the late-time Integrated Sachs-Wolfe (ISW) 
    effect. The fluctuations at high $\ell$ are attributed to the diminished CMB 
    lensing effect. }
\end{figure}

We demonstrate in Fig.~\ref{fig4} the impact on the CMB temperature power spectrum. 
Relative to the $\Lambda$CDM model, the thermal friction model exhibits a pronounced 
increase in power at low $\ell$, which arises from the late-time Integrated Sachs-Wolfe 
(ISW) effect originating from the photon passing through a time-dependent gravitational 
potential. 

In the DADM model, due to the energy transfer from axion dark matter to dark radiation 
caused by thermal friction, the evolution of dark matter is modified, 
resulting in an earlier equality epoch of matter-dark energy. Consequently, this leads 
to a heightened late-time ISW effect. 

Furthermore, the transition from dark matter to dark radiation in the DADM model leads 
to a reduction in CMB lensing. As a result, a decreased peak-smearing can be observed at 
high $\ell$, resulting in a coherent trend of elevated peaks and lowered troughs 
compared to the $\Lambda$CDM model.

\begin{figure}
    \includegraphics[width=\columnwidth]{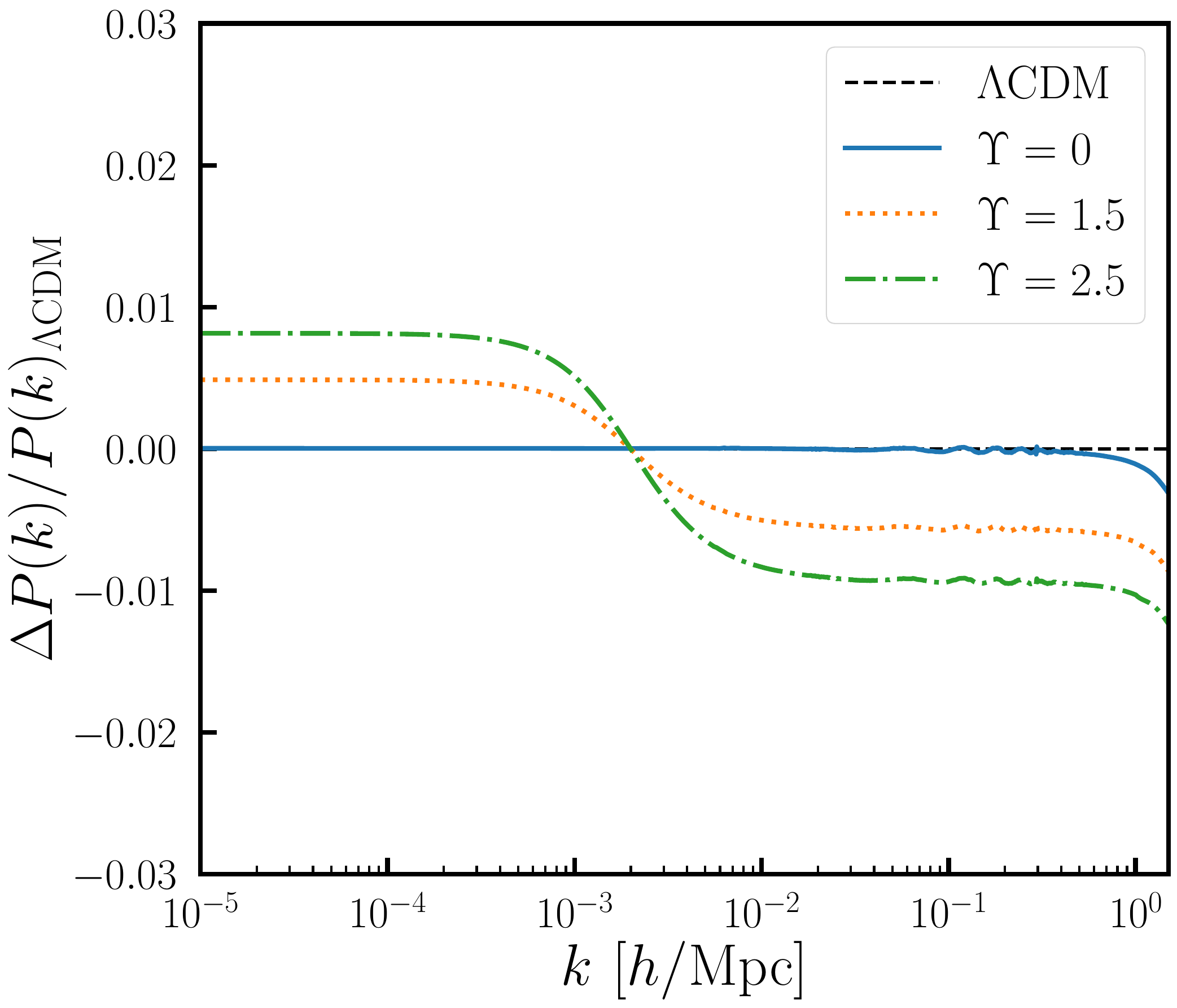}
    \caption{\label{fig5}The results of the linear matter power spectrum relative to 
    the $\Lambda$CDM model are examined under different thermal friction coefficients. 
    In the DADM model, the decrease in the matter density fraction $\Omega_m$ leads to 
    the suppression of the $P(k)$ spectrum. Additionally, the condensation 
    effect of axion dark matter on small scales further contributes to the suppression 
    of the matter power spectrum. }
\end{figure}

Figure~\ref{fig5} illustrates the results of the linear matter power spectrum relative to 
the $\Lambda$CDM model under different thermal friction coefficients. The dissipative 
nature of axion dark matter leads to a decrease in the matter density fraction $\Omega_m$, 
resulting in the suppression of power. This characteristic becomes more 
prominent with an increase in the thermal friction coefficient. 
Due to fixing the energy density fraction of axion dark matter to be equal to that of 
cold dark matter in the $\Lambda$CDM model at the epoch of matter-radiation equality, 
the transition of the power spectrum is correlated with $k_\mathrm{eq}$.

On the smallest scale, the power spectrum of the DADM model exhibits additional reduction 
compared to the $\Lambda$CDM model, as evident from the curve corresponding to the blue 
solid line representing a thermal friction coefficient of 0. This phenomenon arises from 
the inherent properties of axion dark matter, specifically the condensation effect on 
small scales, which further suppresses the matter power spectrum.

\section{\label{sec4}MCMC Analyses}
We employed \texttt{Cobaya} \cite{Torrado_2021} to perform the Markov Chain Monte Carlo 
(MCMC) analysis for obtaining the posterior distribution of model parameters, and 
assessed the convergence using the Gelman-Rubin criterion, $R-1 < 0.01$ 
\cite{Gelman1992InferenceFI}. The MCMC chain was analyzed using
\texttt{GetDist} \cite{lewis2019getdist}.

\subsection{Datasets}
We conducted our analysis using the following cosmological datasets, 
\begin{itemize}
    \item[1.] \textbf{CMB}: The temperature and polarization power spectra from the 
    \textit{Planck} 2018 dataset, encompassing both the low-$\ell$ and high-$\ell$ 
    observations, as well as the CMB lensing power spectrum
    \cite{planck2020,osti_1676388,osti_1775409}.
    \item[2.] \textbf{BAO}: The measurements from the BOSS-DR12 sample, which include 
    the combined LOWZ and CMASS galaxy samples \cite{Alam_2017,Buen_Abad_2018}, 
    as well as the small-$z$ measurements obtained from the 6dFGS and the SDSS DR7 
    \cite{19250.x,stv154}.
    \item[3.] \textbf{Supernovae}: The Pantheon supernovae sample, consists of the 
    relative luminosity distances of 1048 Type Ia supernovae, covering a redshift 
    range from 0.01 to 2.3 \cite{Scolnic_2018}.
\end{itemize}  

By combining data from CMB and BAO, we can conduct acoustic horizon measurements at 
various redshifts, thereby overcoming geometric degeneracies and providing constraints 
on the physical processes occurring between recombination and the redshift at which BAO 
is measured. Moreover, the supernova data obtained from the Pantheon sample plays a 
crucial role in tightly constraining late-time new physic within its specific range of 
measured redshifts. 

\begin{itemize}
    \item[4.] \textbf{SH0ES}: The cosmic distance ladder measurement from SH0ES with 
    $H_0 = 73.04 \pm 1.04$ km/s/Mpc \cite{Riess2021ACM}.
    \item[5.] \textbf{DES}: Dark Energy Survey Year-3 weak lensing and galaxy 
    clustering data, the application of Gaussian constraints on $S_8$ yields a result 
    of $0.776 \pm 0.017$ \cite{PRD.105.023520}.
\end{itemize}

We utilize the $H_0$ measurements from SH0ES to mitigate the impact of prior volume 
effects \cite{PRD.103.123542}. This allows us to evaluate the effectiveness of our 
novel model in addressing the discordance between local $H_0$ measurements and CMB 
inference results. Furthermore, we integrate the $S_8$ data from DES to examine 
the model's capability in alleviating the large-scale structure tension.

\begin{itemize}
    \item[6.] \textbf{Lyman-$\alpha$}: 1D power spectrum of Lyman-$\alpha$ forest flux 
    from quasar samples of SDSS DR14 BOSS and eBOSS surveys \cite{Chabanier_2019}. 
    We employ a compressed version of the Gaussian prior form of the likelihood, which 
    provides estimates for the amplitude and slope of the power 
    spectrum at the pivot redshift of $z_p$ = 3 and wavenumber of $k_p$ = 0.009 s/km 
    $\sim$ 1 h/Mpc \cite{goldstein2023canonical,he2023selfinteracting}.
\end{itemize}

In addition to the commonly used datasets mentioned above, we introduced the Lyman-$\alpha$ 
data. We aim to explore the impact of small-scale measurement on the thermal friction model, 
as axion dark matter exhibits unique features on very small scales.

\subsection{Results}
The mean and 1-$\sigma$ parameter results for $\Lambda$CDM, DADM models, constrained by 
the combined dataset including CMB, BAO, SNIa, SH0ES, and DES, are presented in 
Tab.~\ref{tab1}. Meanwhile, Tab.~\ref{tab2} exhibits the $\chi^2$ statistical values.
\begin{table}
    \caption{\label{tab1}The table presents the mean and 1-$\sigma$ 
    marginalized constraints for $\Lambda$CDM, DADM models using a 
    combined data comprising CMB, BAO, SNIa, SH0ES, and $S_8$ from DES. }
    \renewcommand{\arraystretch}{1.2}
\resizebox{\columnwidth}{!}{
    \begin{tabular} { l  c  c}

       Model  &  $\Lambda$CDM  & DADM\\
       \hline
       \hline
       {\boldmath$\ln(10^{10} A_\mathrm{s})$} & 
       $3.049^{+0.013}_{-0.015}   $&
       $3.054\pm 0.015   $\\
       
       {\boldmath$n_\mathrm{s}   $} & 
       $0.9705\pm 0.0036          $&
       $0.9715\pm 0.0037          $\\
       
       {\boldmath$\tau_\mathrm{reio}$} & 
       $0.0595^{+0.0067}_{-0.0078}$&
       $0.0608^{+0.0072}_{-0.0080}$\\
       
       {\boldmath$H_0            $} & 
       $68.63\pm 0.36             $&
       $68.60\pm 0.40             $\\
       
       {\boldmath$\omega_\mathrm{b}$} & 
       $0.02259\pm 0.00013        $&
       $0.02260\pm 0.00013        $\\
       
       {\boldmath$\omega_\mathrm{dm+dr}$} & 
       $0.11728\pm 0.00077        $&
       $0.11620^{+0.0012}_{-0.00097}$\\

       {\boldmath$\log_{10}m_{\chi}$} & 
       $-             $&
       $< -21.8    $\\     
       
       {\boldmath$\Upsilon$} & 
       $-             $&
       $< 2.42 $\\        
       
       $10^{-9}A_\mathrm{s}    $ & 
       $2.110^{+0.028}_{-0.032}$&
       $2.119\pm 0.033$\\
       
       $\Omega_\mathrm{m}         $ & 
       $0.2984\pm 0.0045          $&
       $0.2926^{+0.0066}_{-0.0056}$\\
       
       $\sigma_8                  $ & 
       $0.8045\pm 0.0056          $&
       $0.8048\pm 0.0059          $\\
       
       $S_8                       $ & 
       $0.8023\pm 0.0085          $&
       $0.795\pm 0.011$\\
       \hline 
       $\Delta\chi_\mathrm{tot}^2 $ & 
       $ - $&
       $-2.60$\\      
       \hline 
    \end{tabular}
}
\end{table}

We find that the DADM model and the $\Lambda$CDM 
model yield similar constraints on the value of $H_0$, indicating that the ability of 
the thermal friction model to resolve the Hubble tension can be considered negligible. 

However, the $S_8$ values obtained from the DADM model is 
smaller than that of the $\Lambda$CDM model, the value of $S_8$ at 68\% confidence 
level for the DADM model is found to be $0.795\pm 0.011$, whereas the result for the 
$\Lambda$CDM model is $0.8023\pm 0.0085$, suggesting that the DADM model can 
alleviate the $S_8$ tension. 

\begin{figure}
    \includegraphics[width=\columnwidth]{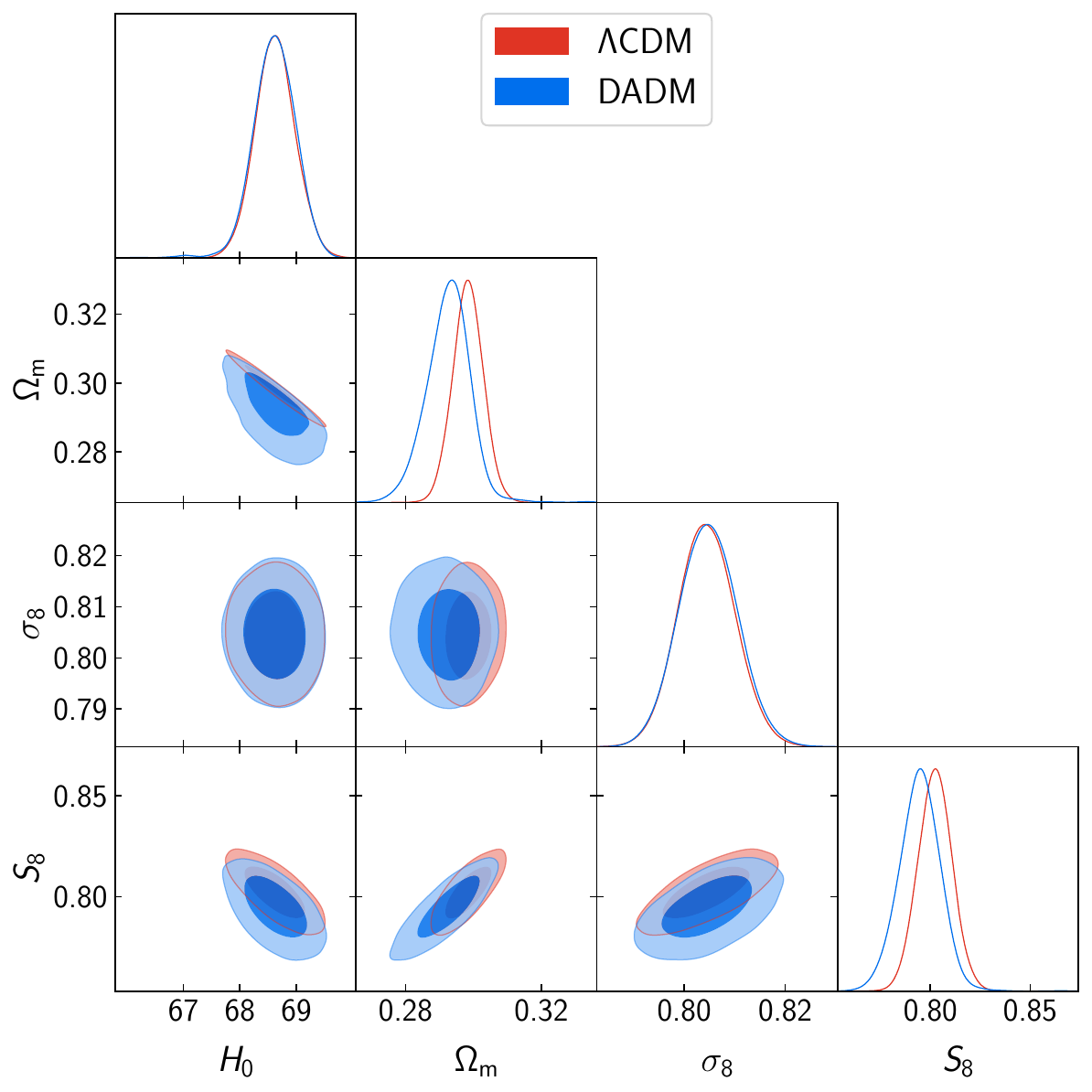}
    \caption{\label{fig6}The posterior distributions of selected parameters for the 
    DADM and $\Lambda$CDM models using a combined data comprising CMB, BAO, SNIa, SH0ES, 
    and $S_8$ from DES are shown. In the DADM model, dark matter undergoes 
    energy transfer to dark radiation through thermal friction, resulting in a lower 
    matter energy density fraction $\Omega_{m}$ and consequently a smaller 
    value of $S_8$.}
\end{figure}

This effect is primarily attributed to the transfer of 
energy from dark matter to dark radiation in the new model, resulting in a lower matter 
energy density fraction $\Omega_{m}$. This characteristic can be more clearly 
observed from the posterior distributions of the model parameters for both models, as 
depicted in Fig.~\ref{fig6}.

From Tab.~\ref{tab2}, it is observable that the performance of the DADM model and the 
$\Lambda$CDM model 
in fitting the CMB and SNIa data is comparable. While the $\Lambda$CDM model fits better 
with the BAO data, the DADM model matches more closely with the results from SH0ES and 
DES. Ultimately, the DADM model yields a smaller $\chi^2_\mathrm{tot}$ of -2.60 compared 
to the $\Lambda$CDM model. 
\begin{table}
    \caption{\label{tab2}The $\chi^2$ statistical values for fitting a combined dataset 
    including CMB, BAO, SNIa, SH0ES, and DES}
    \renewcommand{\arraystretch}{1.2}
\resizebox{.8\columnwidth}{!}{
    \begin{tabular} { l  c  c}

       Datasets  &  $\Lambda$CDM  & DADM\\
       \hline
       \hline
       CMB: & & \\      
       \quad \textit{Planck} 2018 low-$\ell$ TT &22.04 &22.67 \\        
       \quad \textit{Planck} 2018 low-$\ell$ EE &396.10 &397.33 \\        
       \quad \textit{Planck} 2018 high-$\ell$\\ 
       \qquad TT+TE+EE &2352.66 &2351.20 \\  
       LSS: & & \\      
       \quad \textit{Planck} CMB lensing &10.17 &9.57 \\             
       \quad BAO (6dF) &0.0017 &0.0145 \\             
       \quad BAO (DR7 MGS) &1.86 &2.11 \\             
       \quad BAO (DR12 BOSS) &5.858 &6.088 \\         
       SNIa (Pantheon) &1034.76 &1034.73 \\    
       SH0ES &20.55 &18.77 \\    
       DES &3.49 &2.40 \\    
       \hline 
       $\Delta\chi^2_\mathrm{CMB}$ &- &-0.20 \\             
       $\Delta\chi^2_\mathrm{BAO}$ &- &0.49 \\             
       $\Delta\chi^2_\mathrm{SNIa}$ &- &-0.03 \\                
       $\Delta\chi^2_\mathrm{SH0ES}$ &- &-1.78 \\                
       $\Delta\chi^2_\mathrm{DES}$ &- &-1.09 \\                
       \hline
       $\Delta\chi^2_\mathrm{tot}$ &- &-2.60 \\      
       \hline         
    \end{tabular}
}
\end{table}

We obtain an upper limit for the thermal friction 
coefficient $\Upsilon$, with a maximum mean value still below 2.42 km/s/Mpc, 
which is consistent with our assumption of weak thermal friction. The non-zero thermal 
friction coefficient indicates the transfer of energy from axion dark matter to dark 
radiation, thereby contributing to alleviating the $S_8$ tension. 

Moreover, we find that the mass of axion dark matter cannot be constrained when 
utilizing the aforementioned five datasets, as the \texttt{GetDist} analysis chain 
yielded erroneous result. Therefore, we decided to fix the mass of axion dark matter 
at $10^{-22}$ eV and investigate the impact of small-scale observation on the constraint results.

After incorporating Lyman-$\alpha$ data, we re-constrained the DADM model, and the 
constraint results from different combinations of datasets are presented in 
Tab.~\ref{tab3}. The combination of CMB, BAO, SNIa and Lyman-$\alpha$ 
is referred to as the ``Baseline'' in our analysis. 
\begin{table*}
    \caption{\label{tab3}The mean and 1-$\sigma$ marginalized 
    constraints of the cosmological parameters in the DADM model based on different 
    combinations of datasets, where we fixed the mass of axion dark matter to be 
    $10^{-22}$ eV. In this context, ``Baseline'' refers to the combination 
    of CMB, BAO, SNIa, and Lyman-$\alpha$.}
    \renewcommand{\arraystretch}{1.25}
    \tabcolsep=0.25cm
\resizebox{\linewidth}{!}{
    \begin{tabular} { l  c  c  c  c}

        Parameters  &Baseline&Baseline+SH0ES&Baseline+DES&Baseline+SH0E+DES\\
       \hline               
       \hline
       {\boldmath$\ln(10^{10} A_\mathrm{s})$} 
       & $3.057^{+0.014}_{-0.016}   $
       & $3.064^{+0.014}_{-0.016}   $
       & $3.055^{+0.013}_{-0.015}            $
       & $3.062^{+0.014}_{-0.016}            $\\
   
       {\boldmath$n_\mathrm{s}   $} 
       & $0.9601\pm 0.0037          $
       & $0.9641\pm 0.0036          $
       & $0.9613\pm 0.0036          $
       & $0.9646\pm 0.0038          $\\
       
       {\boldmath$\tau_\mathrm{reio}$} 
       & $0.0603^{+0.0069}_{-0.0079}          $
       & $0.0648^{+0.0072}_{-0.0083}          $
       & $0.0598^{+0.0067}_{-0.0077}          $
       & $0.0642\pm 0.0079                    $\\
       
       {\boldmath$H_0            $} 
       & $67.91\pm 0.46             $
       & $68.60^{+0.42}_{-0.37}             $
       & $68.22\pm 0.39             $
       & $68.76^{+0.39}_{-0.35}             $\\
       
       {\boldmath$\omega_\mathrm{b}$} 
       & $0.02247\pm 0.00013        $
       & $0.02262\pm 0.00014        $
       & $0.02251\pm 0.00014        $
       & $0.02263\pm 0.00014        $\\
       
       {\boldmath$\omega_\mathrm{dm+dr}$} 
       & $0.1178^{+0.0013}_{-0.0011}$
       & $0.1163^{+0.0013}_{-0.0011}$
       & $0.1166^{+0.0014}_{-0.0011}$
       & $0.1157^{+0.0012}_{-0.0010}$\\
       
       {\boldmath$\Upsilon       $} 
       & $< 1.81                    $
       & $< 2.06                    $
       & $< 3.18                    $
       & $< 2.74                    $\\
       
       $\Omega_\mathrm{m}         $ 
       & $0.3027^{+0.0073}_{-0.0066}$
       & $0.2934^{+0.0073}_{-0.0063}$
       & $0.2954^{+0.0077}_{-0.0062}$
       & $0.2899^{+0.0066}_{-0.0059}$\\
       
       $\sigma_8                  $ 
       & $0.8084^{+0.0057}_{-0.0065}          $
       & $0.8068\pm 0.0063          $
       & $0.8047\pm 0.0055          $
       & $0.8046\pm 0.0058          $\\
       
       $S_8                       $ 
       & $0.812\pm 0.014            $
       & $0.798\pm 0.013   $
       & $0.799^{+0.011}_{-0.010}  $
       & $0.791\pm 0.011       $\\       
       \hline         
    \end{tabular}
}
\end{table*}

We find that 
the thermal friction coefficient is still small, with its maximum result being less 
than 2.74 km/s/Mpc. 
Additionally, some slight improvements are observed, yielding values of $H_0$ and $S_8$ as 
$68.76^{+0.39}_{-0.35}$ km/s/Mpc and $0.791\pm 0.011$ at a 68\% confidence level, respectively. 

\begin{figure}
    \includegraphics[width=\columnwidth]{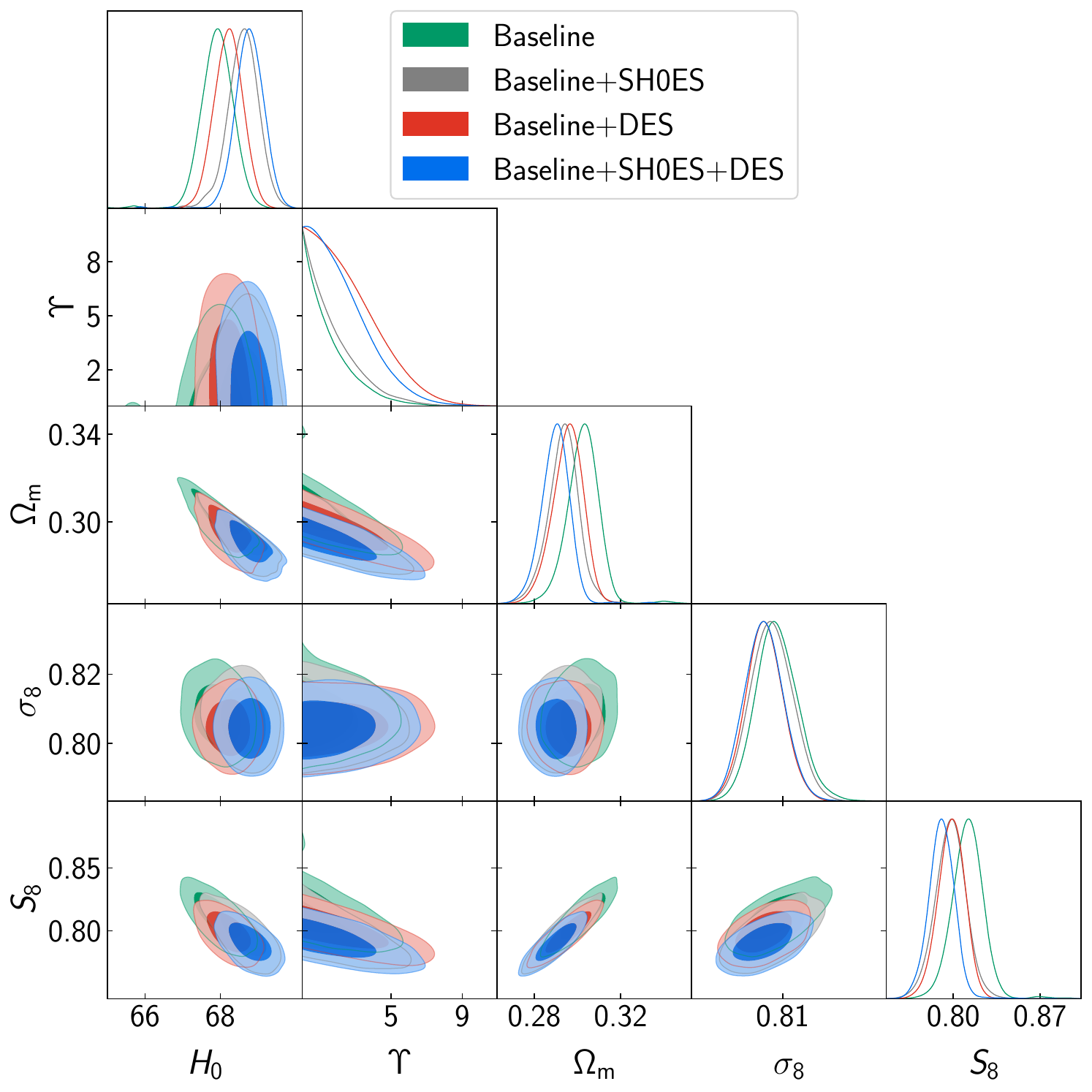}
    \caption{\label{fig7}Constraint results on the DADM model from different 
    combinations of datasets are presented. The green contour represents the outcomes 
    obtained from the baseline dataset, including CMB, BAO, SNIa and Lyman-$\alpha$. 
    The other contours correspond to combinations of the baseline dataset with SH0ES and 
    DES data. }
\end{figure}

Figure~\ref{fig7} shows the constraint results of the DADM model for different 
combinations of datasets. The complete posterior distribution, including all 
parameters, is illustrated in Fig.~\ref{fig8} in the Appendix. 

The green contour represents the results obtained from the 
baseline dataset, which includes CMB, BAO, SNIa and Lyman-$\alpha$. The gray contour 
corresponds to the inclusion of the SH0ES data in addition to the baseline dataset. 
The red contour represents the results obtained by adding the $S_8$ measurement from 
DES data. Lastly, the blue contour indicates the results obtained from the combination 
of all datasets. 

It is noteworthy that when adding the SH0ES and DES data to the baseline dataset, 
the resulting constraints on $S_8$ are remarkably close, which can be explained. The 
effect of SH0ES data is to increase $H_0$, leading to a decrease in $\Omega_m$ to 
maintain consistency with other data. On the other hand, DES data has the effect of 
reducing $\sigma_8$ without altering $\Omega_m$. Since $S_8\equiv\sigma_8\sqrt{(\Omega_m/0.3)}$, 
despite the higher value of $\sigma_8$ from SH0ES data, the final results of $S_8$ 
obtained from both combinations of datasets are similar.

\section{\label{sec5}Conclusion}
In this paper, we investigate the dissipative axion dark matter (DADM) model as a means 
to alleviate cosmological tensions. In this novel model, we replace the cold dark matter 
with axion dark matter and introduce a coupling between the 
dark gauge boson and axion. This coupling leads to additional thermal friction 
experienced by the axion dark matter, resulting in energy injection into the dark 
radiation component.

The interaction between dark matter and dark radiation exhibits similarities to decaying 
dark matter, yet possesses distinct characteristics. We derived the evolving equations 
for interacting dark matter and dark radiation from a scalar field theory framework. 
Notably, the self-interaction of dark radiation generated by thermal friction is 
significantly different from the free-streaming dark radiation resulting from decaying 
dark matter.

We performed the MCMC analysis comparing the new model 
with the $\Lambda$CDM model using commonly used cosmological data, including CMB, BAO, 
SNIa, SH0ES, and $S_8$ from DES.
The results indicate that the constraints on 
$H_0$ obtained from the DADM model are in close agreement with those of the $\Lambda$CDM 
model. However, the value of $S_8$ in the DADM model is smaller than 
that in the $\Lambda$CDM model. The value of $S_8$ for the DADM model is found to be 
$0.795\pm 0.011$ at a 68\% confidence level, whereas the result for the $\Lambda$CDM model 
is $0.8023\pm 0.0085$. This suggests that the new model can partially alleviate the $S_8$ 
tension but its ability to address the Hubble tension can be ignored.

According to the $\chi^2$ statistic, we find that the $\Lambda$CDM model provides a 
better fit to the BAO data, while the 
DADM model is more consistent with the SH0ES and DES data. Besides, there is little 
difference in the agreement between the two models and CMB and SNIa data. Nonetheless, 
the DADM model has a greater advantage in fitting the SH0ES and DES data, resulting in 
a smaller $\chi^2_\mathrm{tot}$ of -2.60 compared to the $\Lambda$CDM model.

In addition, we found that the five aforementioned datasets are unable to constrain the 
mass of axion dark matter. Therefore, we fixed the mass of axion dark matter to be $10^{-22}$ eV, 
and included small-scale data to explore its impact on the model parameters. 
After incorporating the Lyman-$\alpha$ data and performing the new round of MCMC analysis, 
we found that the results show slight improvements, with values of $H_0$ and $S_8$ being 
$68.76^{+0.39}_{-0.35}$ km/s/Mpc and $0.791\pm 0.011$ at a 68\% confidence level, respectively.

The axion dark matter thermal friction model still fails to fully resolve the 
cosmological tensions. Given that in this paper, we only consider the simplest 
scenario involving a constant thermal friction coefficient and the assumption of weak 
thermal friction, more complex models can be considered to address the Hubble and 
large-scale structure tensions.

\begin{acknowledgments}
    This work is supported in part by National Natural Science Foundation of China 
    under Grant No.12075042, Grant No.11675032 (People's Republic of China).
\end{acknowledgments}

\appendix*
\section{The full MCMC posteriors}
\begin{figure*}
    \includegraphics[width=\linewidth]{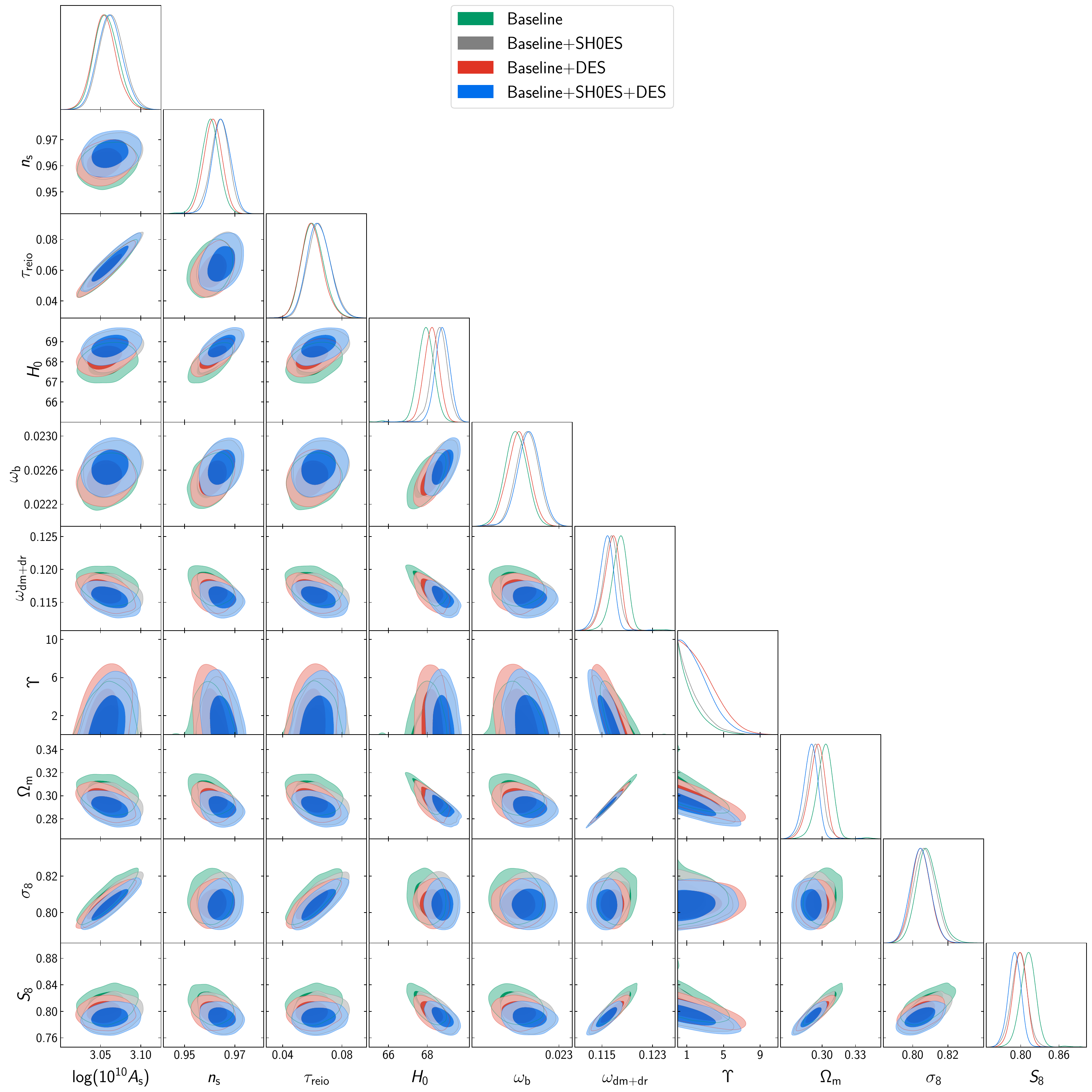}
    \caption{\label{fig8}The complete Markov Chain Monte Carlo posterior of the DADM 
    model parameters obtained using different combinations of datasets is presented. 
    Various data, including CMB, BAO, SNIa, Lyman-$\alpha$, SH0ES, and $S_8$ 
    from DES-Y3, are utilized in our analysis.}
\end{figure*}

\bibliography{dadm}

\begin{thebibliography}{76}%
\makeatletter
\providecommand \@ifxundefined [1]{%
 \@ifx{#1\undefined}
}%
\providecommand \@ifnum [1]{%
 \ifnum #1\expandafter \@firstoftwo
 \else \expandafter \@secondoftwo
 \fi
}%
\providecommand \@ifx [1]{%
 \ifx #1\expandafter \@firstoftwo
 \else \expandafter \@secondoftwo
 \fi
}%
\providecommand \natexlab [1]{#1}%
\providecommand \enquote  [1]{``#1''}%
\providecommand \bibnamefont  [1]{#1}%
\providecommand \bibfnamefont [1]{#1}%
\providecommand \citenamefont [1]{#1}%
\providecommand \href@noop [0]{\@secondoftwo}%
\providecommand \href [0]{\begingroup \@sanitize@url \@href}%
\providecommand \@href[1]{\@@startlink{#1}\@@href}%
\providecommand \@@href[1]{\endgroup#1\@@endlink}%
\providecommand \@sanitize@url [0]{\catcode `\\12\catcode `\$12\catcode `\&12\catcode `\#12\catcode `\^12\catcode `\_12\catcode `\%12\relax}%
\providecommand \@@startlink[1]{}%
\providecommand \@@endlink[0]{}%
\providecommand \url  [0]{\begingroup\@sanitize@url \@url }%
\providecommand \@url [1]{\endgroup\@href {#1}{\urlprefix }}%
\providecommand \urlprefix  [0]{URL }%
\providecommand \Eprint [0]{\href }%
\providecommand \doibase [0]{https://doi.org/}%
\providecommand \selectlanguage [0]{\@gobble}%
\providecommand \bibinfo  [0]{\@secondoftwo}%
\providecommand \bibfield  [0]{\@secondoftwo}%
\providecommand \translation [1]{[#1]}%
\providecommand \BibitemOpen [0]{}%
\providecommand \bibitemStop [0]{}%
\providecommand \bibitemNoStop [0]{.\EOS\space}%
\providecommand \EOS [0]{\spacefactor3000\relax}%
\providecommand \BibitemShut  [1]{\csname bibitem#1\endcsname}%
\let\auto@bib@innerbib\@empty
\bibitem [{\citenamefont {Verde}\ \emph {et~al.}(2019)\citenamefont {Verde}, \citenamefont {Treu},\ and\ \citenamefont {Riess}}]{Verde_2019}%
  \BibitemOpen
  \bibfield  {author} {\bibinfo {author} {\bibfnamefont {L.}~\bibnamefont {Verde}}, \bibinfo {author} {\bibfnamefont {T.}~\bibnamefont {Treu}},\ and\ \bibinfo {author} {\bibfnamefont {A.~G.}\ \bibnamefont {Riess}},\ }\bibfield  {title} {\bibinfo {title} {Tensions between the early and late universe},\ }\href {https://doi.org/10.1038/s41550-019-0902-0} {\bibfield  {journal} {\bibinfo  {journal} {Nature Astronomy}\ }\textbf {\bibinfo {volume} {3}},\ \bibinfo {pages} {891} (\bibinfo {year} {2019})}\BibitemShut {NoStop}%
\bibitem [{\citenamefont {Freedman}(2017)}]{freedman2017}%
  \BibitemOpen
  \bibfield  {author} {\bibinfo {author} {\bibfnamefont {W.~L.}\ \bibnamefont {Freedman}},\ }\bibfield  {title} {\bibinfo {title} {Cosmology at at crossroads: Tension with the hubble constant},\ }\bibfield  {journal} {\bibinfo  {journal} {Nature Astronomy}\ }\href {https://doi.org/10.1038/s41550-017-0121} {10.1038/s41550-017-0121} (\bibinfo {year} {2017})\BibitemShut {NoStop}%
\bibitem [{\citenamefont {{Planck Collaboration}}\ \emph {et~al.}(2020)\citenamefont {{Planck Collaboration}}, \citenamefont {{Aghanim, N.}}, \citenamefont {{Akrami, Y.}} \emph {et~al.}}]{planck2020}%
  \BibitemOpen
  \bibfield  {author} {\bibinfo {author} {\bibnamefont {{Planck Collaboration}}}, \bibinfo {author} {\bibnamefont {{Aghanim, N.}}}, \bibinfo {author} {\bibnamefont {{Akrami, Y.}}}, \emph {et~al.},\ }\bibfield  {title} {\bibinfo {title} {{Planck 2018 results - VI. Cosmological parameters}},\ }\href {https://doi.org/10.1051/0004-6361/201833910} {\bibfield  {journal} {\bibinfo  {journal} {Astronomy and Astrophysics}\ }\textbf {\bibinfo {volume} {641}},\ \bibinfo {pages} {A6} (\bibinfo {year} {2020})}\BibitemShut {NoStop}%
\bibitem [{\citenamefont {Riess}\ \emph {et~al.}(2021)\citenamefont {Riess}, \citenamefont {Yuan}, \citenamefont {Macri} \emph {et~al.}}]{Riess2021ACM}%
  \BibitemOpen
  \bibfield  {author} {\bibinfo {author} {\bibfnamefont {A.~G.}\ \bibnamefont {Riess}}, \bibinfo {author} {\bibfnamefont {W.}~\bibnamefont {Yuan}}, \bibinfo {author} {\bibfnamefont {L.~M.}\ \bibnamefont {Macri}}, \emph {et~al.},\ }\bibfield  {title} {\bibinfo {title} {{A Comprehensive Measurement of the Local Value of the Hubble Constant with 1 km/s/Mpc Uncertainty from the Hubble Space Telescope and the SH0ES Team}}\ }(\bibinfo {year} {2021})\ \Eprint {https://arxiv.org/abs/2112.04510} {arXiv:2112.04510 [astro-ph.CO]} \BibitemShut {NoStop}%
\bibitem [{\citenamefont {Abbott}\ \emph {et~al.}(2022)\citenamefont {Abbott}, \citenamefont {Aguena}, \citenamefont {Alarcon} \emph {et~al.}}]{PRD.105.023520}%
  \BibitemOpen
  \bibfield  {author} {\bibinfo {author} {\bibfnamefont {T.~M.~C.}\ \bibnamefont {Abbott}}, \bibinfo {author} {\bibfnamefont {M.}~\bibnamefont {Aguena}}, \bibinfo {author} {\bibfnamefont {A.}~\bibnamefont {Alarcon}}, \emph {et~al.},\ }\bibfield  {title} {\bibinfo {title} {Dark energy survey year 3 results: Cosmological constraints from galaxy clustering and weak lensing},\ }\href {https://doi.org/10.1103/PhysRevD.105.023520} {\bibfield  {journal} {\bibinfo  {journal} {Phys. Rev. D}\ }\textbf {\bibinfo {volume} {105}},\ \bibinfo {pages} {023520} (\bibinfo {year} {2022})}\BibitemShut {NoStop}%
\bibitem [{\citenamefont {Li}\ and\ \citenamefont {Shafieloo}(2019)}]{Li_2019}%
  \BibitemOpen
  \bibfield  {author} {\bibinfo {author} {\bibfnamefont {X.}~\bibnamefont {Li}}\ and\ \bibinfo {author} {\bibfnamefont {A.}~\bibnamefont {Shafieloo}},\ }\bibfield  {title} {\bibinfo {title} {A simple phenomenological emergent dark energy model can resolve the hubble tension},\ }\href {https://doi.org/10.3847/2041-8213/ab3e09} {\bibfield  {journal} {\bibinfo  {journal} {The Astrophysical Journal Letters}\ }\textbf {\bibinfo {volume} {883}},\ \bibinfo {pages} {L3} (\bibinfo {year} {2019})}\BibitemShut {NoStop}%
\bibitem [{\citenamefont {Zhou}\ \emph {et~al.}(2022{\natexlab{a}})\citenamefont {Zhou}, \citenamefont {Liu}, \citenamefont {Mu},\ and\ \citenamefont {Xu}}]{Zhouzh}%
  \BibitemOpen
  \bibfield  {author} {\bibinfo {author} {\bibfnamefont {Z.}~\bibnamefont {Zhou}}, \bibinfo {author} {\bibfnamefont {G.}~\bibnamefont {Liu}}, \bibinfo {author} {\bibfnamefont {Y.}~\bibnamefont {Mu}},\ and\ \bibinfo {author} {\bibfnamefont {L.}~\bibnamefont {Xu}},\ }\bibfield  {title} {\bibinfo {title} {{Can phantom transition at z ~ 1 restore the Cosmic concordance?}},\ }\href {https://doi.org/10.1093/mnras/stac053} {\bibfield  {journal} {\bibinfo  {journal} {Monthly Notices of the Royal Astronomical Society}\ }\textbf {\bibinfo {volume} {511}},\ \bibinfo {pages} {595} (\bibinfo {year} {2022}{\natexlab{a}})}\BibitemShut {NoStop}%
\bibitem [{\citenamefont {Guo}\ \emph {et~al.}(2019)\citenamefont {Guo}, \citenamefont {Zhang},\ and\ \citenamefont {Zhang}}]{Guo_2019}%
  \BibitemOpen
  \bibfield  {author} {\bibinfo {author} {\bibfnamefont {R.-Y.}\ \bibnamefont {Guo}}, \bibinfo {author} {\bibfnamefont {J.-F.}\ \bibnamefont {Zhang}},\ and\ \bibinfo {author} {\bibfnamefont {X.}~\bibnamefont {Zhang}},\ }\bibfield  {title} {\bibinfo {title} {Can the h0 tension be resolved in extensions to lcdm cosmology?},\ }\href {https://doi.org/10.1088/1475-7516/2019/02/054} {\bibfield  {journal} {\bibinfo  {journal} {Journal of Cosmology and Astroparticle Physics}\ }\textbf {\bibinfo {volume} {2019}}\bibinfo  {number} { (02)},\ \bibinfo {pages} {054}}\BibitemShut {NoStop}%
\bibitem [{\citenamefont {Di~Valentino}\ \emph {et~al.}(2018)\citenamefont {Di~Valentino}, \citenamefont {Linder},\ and\ \citenamefont {Melchiorri}}]{PRD.97.043528}%
  \BibitemOpen
\bibfield  {number} {  }\bibfield  {author} {\bibinfo {author} {\bibfnamefont {E.}~\bibnamefont {Di~Valentino}}, \bibinfo {author} {\bibfnamefont {E.~V.}\ \bibnamefont {Linder}},\ and\ \bibinfo {author} {\bibfnamefont {A.}~\bibnamefont {Melchiorri}},\ }\bibfield  {title} {\bibinfo {title} {Vacuum phase transition solves the ${H}_{0}$ tension},\ }\href {https://doi.org/10.1103/PhysRevD.97.043528} {\bibfield  {journal} {\bibinfo  {journal} {Phys. Rev. D}\ }\textbf {\bibinfo {volume} {97}},\ \bibinfo {pages} {043528} (\bibinfo {year} {2018})}\BibitemShut {NoStop}%
\bibitem [{\citenamefont {Banihashemi}\ \emph {et~al.}(2019)\citenamefont {Banihashemi}, \citenamefont {Khosravi},\ and\ \citenamefont {Shirazi}}]{PRD.99.083509}%
  \BibitemOpen
  \bibfield  {author} {\bibinfo {author} {\bibfnamefont {A.}~\bibnamefont {Banihashemi}}, \bibinfo {author} {\bibfnamefont {N.}~\bibnamefont {Khosravi}},\ and\ \bibinfo {author} {\bibfnamefont {A.~H.}\ \bibnamefont {Shirazi}},\ }\bibfield  {title} {\bibinfo {title} {Ginzburg-landau theory of dark energy: A framework to study both temporal and spatial cosmological tensions simultaneously},\ }\href {https://doi.org/10.1103/PhysRevD.99.083509} {\bibfield  {journal} {\bibinfo  {journal} {Phys. Rev. D}\ }\textbf {\bibinfo {volume} {99}},\ \bibinfo {pages} {083509} (\bibinfo {year} {2019})}\BibitemShut {NoStop}%
\bibitem [{\citenamefont {Banihashemi}\ \emph {et~al.}(2020)\citenamefont {Banihashemi}, \citenamefont {Khosravi},\ and\ \citenamefont {Shirazi}}]{PRD.101.123521}%
  \BibitemOpen
  \bibfield  {author} {\bibinfo {author} {\bibfnamefont {A.}~\bibnamefont {Banihashemi}}, \bibinfo {author} {\bibfnamefont {N.}~\bibnamefont {Khosravi}},\ and\ \bibinfo {author} {\bibfnamefont {A.~H.}\ \bibnamefont {Shirazi}},\ }\bibfield  {title} {\bibinfo {title} {Phase transition in the dark sector as a proposal to lessen cosmological tensions},\ }\href {https://doi.org/10.1103/PhysRevD.101.123521} {\bibfield  {journal} {\bibinfo  {journal} {Phys. Rev. D}\ }\textbf {\bibinfo {volume} {101}},\ \bibinfo {pages} {123521} (\bibinfo {year} {2020})}\BibitemShut {NoStop}%
\bibitem [{\citenamefont {Poulin}\ \emph {et~al.}(2019)\citenamefont {Poulin}, \citenamefont {Smith}, \citenamefont {Karwal},\ and\ \citenamefont {Kamionkowski}}]{PRL.122.221301}%
  \BibitemOpen
  \bibfield  {author} {\bibinfo {author} {\bibfnamefont {V.}~\bibnamefont {Poulin}}, \bibinfo {author} {\bibfnamefont {T.~L.}\ \bibnamefont {Smith}}, \bibinfo {author} {\bibfnamefont {T.}~\bibnamefont {Karwal}},\ and\ \bibinfo {author} {\bibfnamefont {M.}~\bibnamefont {Kamionkowski}},\ }\bibfield  {title} {\bibinfo {title} {Early dark energy can resolve the hubble tension},\ }\href {https://doi.org/10.1103/PhysRevLett.122.221301} {\bibfield  {journal} {\bibinfo  {journal} {Phys. Rev. Lett.}\ }\textbf {\bibinfo {volume} {122}},\ \bibinfo {pages} {221301} (\bibinfo {year} {2019})}\BibitemShut {NoStop}%
\bibitem [{\citenamefont {Smith}\ \emph {et~al.}(2020)\citenamefont {Smith}, \citenamefont {Poulin},\ and\ \citenamefont {Amin}}]{PRD.101.063523}%
  \BibitemOpen
  \bibfield  {author} {\bibinfo {author} {\bibfnamefont {T.~L.}\ \bibnamefont {Smith}}, \bibinfo {author} {\bibfnamefont {V.}~\bibnamefont {Poulin}},\ and\ \bibinfo {author} {\bibfnamefont {M.~A.}\ \bibnamefont {Amin}},\ }\bibfield  {title} {\bibinfo {title} {Oscillating scalar fields and the hubble tension: A resolution with novel signatures},\ }\href {https://doi.org/10.1103/PhysRevD.101.063523} {\bibfield  {journal} {\bibinfo  {journal} {Phys. Rev. D}\ }\textbf {\bibinfo {volume} {101}},\ \bibinfo {pages} {063523} (\bibinfo {year} {2020})}\BibitemShut {NoStop}%
\bibitem [{\citenamefont {Hill}\ \emph {et~al.}(2020)\citenamefont {Hill}, \citenamefont {McDonough}, \citenamefont {Toomey},\ and\ \citenamefont {Alexander}}]{PRD.102.043507}%
  \BibitemOpen
  \bibfield  {author} {\bibinfo {author} {\bibfnamefont {J.~C.}\ \bibnamefont {Hill}}, \bibinfo {author} {\bibfnamefont {E.}~\bibnamefont {McDonough}}, \bibinfo {author} {\bibfnamefont {M.~W.}\ \bibnamefont {Toomey}},\ and\ \bibinfo {author} {\bibfnamefont {S.}~\bibnamefont {Alexander}},\ }\bibfield  {title} {\bibinfo {title} {Early dark energy does not restore cosmological concordance},\ }\href {https://doi.org/10.1103/PhysRevD.102.043507} {\bibfield  {journal} {\bibinfo  {journal} {Phys. Rev. D}\ }\textbf {\bibinfo {volume} {102}},\ \bibinfo {pages} {043507} (\bibinfo {year} {2020})}\BibitemShut {NoStop}%
\bibitem [{\citenamefont {Yang}\ \emph {et~al.}(2018{\natexlab{a}})\citenamefont {Yang}, \citenamefont {Pan}, \citenamefont {Xu},\ and\ \citenamefont {Mota}}]{sty2789}%
  \BibitemOpen
  \bibfield  {author} {\bibinfo {author} {\bibfnamefont {W.}~\bibnamefont {Yang}}, \bibinfo {author} {\bibfnamefont {S.}~\bibnamefont {Pan}}, \bibinfo {author} {\bibfnamefont {L.}~\bibnamefont {Xu}},\ and\ \bibinfo {author} {\bibfnamefont {D.~F.}\ \bibnamefont {Mota}},\ }\bibfield  {title} {\bibinfo {title} {{Effects of anisotropic stress in interacting dark matter-dark energy scenarios}},\ }\href {https://doi.org/10.1093/mnras/sty2789} {\bibfield  {journal} {\bibinfo  {journal} {Monthly Notices of the Royal Astronomical Society}\ }\textbf {\bibinfo {volume} {482}},\ \bibinfo {pages} {1858} (\bibinfo {year} {2018}{\natexlab{a}})}\BibitemShut {NoStop}%
\bibitem [{\citenamefont {Valentino}\ \emph {et~al.}(2020)\citenamefont {Valentino}, \citenamefont {Melchiorri}, \citenamefont {Mena},\ and\ \citenamefont {Vagnozzi}}]{Di_Valentino_2020}%
  \BibitemOpen
  \bibfield  {author} {\bibinfo {author} {\bibfnamefont {E.~D.}\ \bibnamefont {Valentino}}, \bibinfo {author} {\bibfnamefont {A.}~\bibnamefont {Melchiorri}}, \bibinfo {author} {\bibfnamefont {O.}~\bibnamefont {Mena}},\ and\ \bibinfo {author} {\bibfnamefont {S.}~\bibnamefont {Vagnozzi}},\ }\href {https://doi.org/10.1016/j.dark.2020.100666} {\bibfield  {journal} {\bibinfo  {journal} {Physics of the Dark Universe}\ }\textbf {\bibinfo {volume} {30}},\ \bibinfo {pages} {100666} (\bibinfo {year} {2020})}\BibitemShut {NoStop}%
\bibitem [{\citenamefont {Yang}\ \emph {et~al.}(2018{\natexlab{b}})\citenamefont {Yang}, \citenamefont {Pan},\ and\ \citenamefont {Paliathanasis}}]{sty2780}%
  \BibitemOpen
  \bibfield  {author} {\bibinfo {author} {\bibfnamefont {W.}~\bibnamefont {Yang}}, \bibinfo {author} {\bibfnamefont {S.}~\bibnamefont {Pan}},\ and\ \bibinfo {author} {\bibfnamefont {A.}~\bibnamefont {Paliathanasis}},\ }\bibfield  {title} {\bibinfo {title} {{Cosmological constraints on an exponential interaction in the dark sector}},\ }\href {https://doi.org/10.1093/mnras/sty2780} {\bibfield  {journal} {\bibinfo  {journal} {Monthly Notices of the Royal Astronomical Society}\ }\textbf {\bibinfo {volume} {482}},\ \bibinfo {pages} {1007} (\bibinfo {year} {2018}{\natexlab{b}})}\BibitemShut {NoStop}%
\bibitem [{\citenamefont {Liu}\ \emph {et~al.}(2023{\natexlab{a}})\citenamefont {Liu}, \citenamefont {Zhou}, \citenamefont {Mu},\ and\ \citenamefont {Xu}}]{liu2023alleviating}%
  \BibitemOpen
  \bibfield  {author} {\bibinfo {author} {\bibfnamefont {G.}~\bibnamefont {Liu}}, \bibinfo {author} {\bibfnamefont {Z.}~\bibnamefont {Zhou}}, \bibinfo {author} {\bibfnamefont {Y.}~\bibnamefont {Mu}},\ and\ \bibinfo {author} {\bibfnamefont {L.}~\bibnamefont {Xu}},\ }\bibfield  {title} {\bibinfo {title} {Alleviating cosmological tensions with a coupled scalar fields model},\ }\href {https://doi.org/10.1103/PhysRevD.108.083523} {\bibfield  {journal} {\bibinfo  {journal} {Phys. Rev. D}\ }\textbf {\bibinfo {volume} {108}},\ \bibinfo {pages} {083523} (\bibinfo {year} {2023}{\natexlab{a}})}\BibitemShut {NoStop}%
\bibitem [{\citenamefont {Liu}\ \emph {et~al.}(2023{\natexlab{b}})\citenamefont {Liu}, \citenamefont {Zhou}, \citenamefont {Mu},\ and\ \citenamefont {Xu}}]{liu2023kinetically}%
  \BibitemOpen
  \bibfield  {author} {\bibinfo {author} {\bibfnamefont {G.}~\bibnamefont {Liu}}, \bibinfo {author} {\bibfnamefont {Z.}~\bibnamefont {Zhou}}, \bibinfo {author} {\bibfnamefont {Y.}~\bibnamefont {Mu}},\ and\ \bibinfo {author} {\bibfnamefont {L.}~\bibnamefont {Xu}},\ }\href@noop {} {\bibinfo {title} {Kinetically coupled scalar fields model and cosmological tensions}} (\bibinfo {year} {2023}{\natexlab{b}}),\ \Eprint {https://arxiv.org/abs/2308.07069} {arXiv:2308.07069 [astro-ph.CO]} \BibitemShut {NoStop}%
\bibitem [{\citenamefont {Liu}\ \emph {et~al.}(2023{\natexlab{c}})\citenamefont {Liu}, \citenamefont {Gao}, \citenamefont {Han}, \citenamefont {Mu},\ and\ \citenamefont {Xu}}]{liu2023mitigating}%
  \BibitemOpen
  \bibfield  {author} {\bibinfo {author} {\bibfnamefont {G.}~\bibnamefont {Liu}}, \bibinfo {author} {\bibfnamefont {J.}~\bibnamefont {Gao}}, \bibinfo {author} {\bibfnamefont {Y.}~\bibnamefont {Han}}, \bibinfo {author} {\bibfnamefont {Y.}~\bibnamefont {Mu}},\ and\ \bibinfo {author} {\bibfnamefont {L.}~\bibnamefont {Xu}},\ }\href@noop {} {\bibinfo {title} {Mitigating cosmological tensions via momentum-coupled dark sector model}} (\bibinfo {year} {2023}{\natexlab{c}}),\ \Eprint {https://arxiv.org/abs/2310.09798} {arXiv:2310.09798 [astro-ph.CO]} \BibitemShut {NoStop}%
\bibitem [{\citenamefont {Raveri}\ \emph {et~al.}(2017)\citenamefont {Raveri}, \citenamefont {Hu}, \citenamefont {Hoffman},\ and\ \citenamefont {Wang}}]{PRD.96.103501}%
  \BibitemOpen
  \bibfield  {author} {\bibinfo {author} {\bibfnamefont {M.}~\bibnamefont {Raveri}}, \bibinfo {author} {\bibfnamefont {W.}~\bibnamefont {Hu}}, \bibinfo {author} {\bibfnamefont {T.}~\bibnamefont {Hoffman}},\ and\ \bibinfo {author} {\bibfnamefont {L.-T.}\ \bibnamefont {Wang}},\ }\bibfield  {title} {\bibinfo {title} {Partially acoustic dark matter cosmology and cosmological constraints},\ }\href {https://doi.org/10.1103/PhysRevD.96.103501} {\bibfield  {journal} {\bibinfo  {journal} {Phys. Rev. D}\ }\textbf {\bibinfo {volume} {96}},\ \bibinfo {pages} {103501} (\bibinfo {year} {2017})}\BibitemShut {NoStop}%
\bibitem [{\citenamefont {Vattis}\ \emph {et~al.}(2019)\citenamefont {Vattis}, \citenamefont {Koushiappas},\ and\ \citenamefont {Loeb}}]{PRD.99.121302}%
  \BibitemOpen
  \bibfield  {author} {\bibinfo {author} {\bibfnamefont {K.}~\bibnamefont {Vattis}}, \bibinfo {author} {\bibfnamefont {S.~M.}\ \bibnamefont {Koushiappas}},\ and\ \bibinfo {author} {\bibfnamefont {A.}~\bibnamefont {Loeb}},\ }\bibfield  {title} {\bibinfo {title} {Dark matter decaying in the late universe can relieve the ${H}_{0}$ tension},\ }\href {https://doi.org/10.1103/PhysRevD.99.121302} {\bibfield  {journal} {\bibinfo  {journal} {Phys. Rev. D}\ }\textbf {\bibinfo {volume} {99}},\ \bibinfo {pages} {121302} (\bibinfo {year} {2019})}\BibitemShut {NoStop}%
\bibitem [{\citenamefont {Buch}\ \emph {et~al.}(2017)\citenamefont {Buch}, \citenamefont {Ralegankar},\ and\ \citenamefont {Rentala}}]{Buch_2017}%
  \BibitemOpen
  \bibfield  {author} {\bibinfo {author} {\bibfnamefont {J.}~\bibnamefont {Buch}}, \bibinfo {author} {\bibfnamefont {P.}~\bibnamefont {Ralegankar}},\ and\ \bibinfo {author} {\bibfnamefont {V.}~\bibnamefont {Rentala}},\ }\bibfield  {title} {\bibinfo {title} {Late decaying 2-component dark matter scenario as an explanation of the ams-02 positron excess},\ }\href {https://doi.org/10.1088/1475-7516/2017/10/028} {\bibfield  {journal} {\bibinfo  {journal} {Journal of Cosmology and Astroparticle Physics}\ }\textbf {\bibinfo {volume} {2017}}\bibinfo  {number} { (10)},\ \bibinfo {pages} {028}}\BibitemShut {NoStop}%
\bibitem [{\citenamefont {Bringmann}\ \emph {et~al.}(2018)\citenamefont {Bringmann}, \citenamefont {Kahlhoefer}, \citenamefont {Schmidt-Hoberg},\ and\ \citenamefont {Walia}}]{PRD.98.023543}%
  \BibitemOpen
\bibfield  {number} {  }\bibfield  {author} {\bibinfo {author} {\bibfnamefont {T.}~\bibnamefont {Bringmann}}, \bibinfo {author} {\bibfnamefont {F.}~\bibnamefont {Kahlhoefer}}, \bibinfo {author} {\bibfnamefont {K.}~\bibnamefont {Schmidt-Hoberg}},\ and\ \bibinfo {author} {\bibfnamefont {P.}~\bibnamefont {Walia}},\ }\bibfield  {title} {\bibinfo {title} {Converting nonrelativistic dark matter to radiation},\ }\href {https://doi.org/10.1103/PhysRevD.98.023543} {\bibfield  {journal} {\bibinfo  {journal} {Phys. Rev. D}\ }\textbf {\bibinfo {volume} {98}},\ \bibinfo {pages} {023543} (\bibinfo {year} {2018})}\BibitemShut {NoStop}%
\bibitem [{\citenamefont {Clark}\ \emph {et~al.}(2021)\citenamefont {Clark}, \citenamefont {Vattis},\ and\ \citenamefont {Koushiappas}}]{PRD.103.043014}%
  \BibitemOpen
  \bibfield  {author} {\bibinfo {author} {\bibfnamefont {S.~J.}\ \bibnamefont {Clark}}, \bibinfo {author} {\bibfnamefont {K.}~\bibnamefont {Vattis}},\ and\ \bibinfo {author} {\bibfnamefont {S.~M.}\ \bibnamefont {Koushiappas}},\ }\bibfield  {title} {\bibinfo {title} {Cosmological constraints on late-universe decaying dark matter as a solution to the ${H}_{0}$ tension},\ }\href {https://doi.org/10.1103/PhysRevD.103.043014} {\bibfield  {journal} {\bibinfo  {journal} {Phys. Rev. D}\ }\textbf {\bibinfo {volume} {103}},\ \bibinfo {pages} {043014} (\bibinfo {year} {2021})}\BibitemShut {NoStop}%
\bibitem [{\citenamefont {Pandey}\ \emph {et~al.}(2020)\citenamefont {Pandey}, \citenamefont {Karwal},\ and\ \citenamefont {Das}}]{Pandey_2020}%
  \BibitemOpen
  \bibfield  {author} {\bibinfo {author} {\bibfnamefont {K.~L.}\ \bibnamefont {Pandey}}, \bibinfo {author} {\bibfnamefont {T.}~\bibnamefont {Karwal}},\ and\ \bibinfo {author} {\bibfnamefont {S.}~\bibnamefont {Das}},\ }\href {https://doi.org/10.1088/1475-7516/2020/07/026} {\bibfield  {journal} {\bibinfo  {journal} {Journal of Cosmology and Astroparticle Physics}\ }\textbf {\bibinfo {volume} {2020}}\bibinfo  {number} { (07)},\ \bibinfo {pages} {026}}\BibitemShut {NoStop}%
\bibitem [{\citenamefont {Alvi}\ \emph {et~al.}(2022)\citenamefont {Alvi}, \citenamefont {Brinckmann}, \citenamefont {Gerbino}, \citenamefont {Lattanzi},\ and\ \citenamefont {Pagano}}]{Alvi_2022}%
  \BibitemOpen
\bibfield  {number} {  }\bibfield  {author} {\bibinfo {author} {\bibfnamefont {S.}~\bibnamefont {Alvi}}, \bibinfo {author} {\bibfnamefont {T.}~\bibnamefont {Brinckmann}}, \bibinfo {author} {\bibfnamefont {M.}~\bibnamefont {Gerbino}}, \bibinfo {author} {\bibfnamefont {M.}~\bibnamefont {Lattanzi}},\ and\ \bibinfo {author} {\bibfnamefont {L.}~\bibnamefont {Pagano}},\ }\bibfield  {title} {\bibinfo {title} {Do you smell something decaying? updated linear constraints on decaying dark matter scenarios},\ }\href {https://doi.org/10.1088/1475-7516/2022/11/015} {\bibfield  {journal} {\bibinfo  {journal} {Journal of Cosmology and Astroparticle Physics}\ }\textbf {\bibinfo {volume} {2022}}\bibinfo  {number} { (11)},\ \bibinfo {pages} {015}}\BibitemShut {NoStop}%
\bibitem [{\citenamefont {McCarthy}\ and\ \citenamefont {Hill}(2023)}]{mccarthy2023converting}%
  \BibitemOpen
\bibfield  {number} {  }\bibfield  {author} {\bibinfo {author} {\bibfnamefont {F.}~\bibnamefont {McCarthy}}\ and\ \bibinfo {author} {\bibfnamefont {J.~C.}\ \bibnamefont {Hill}},\ }\href@noop {} {\bibinfo {title} {Converting dark matter to dark radiation does not solve cosmological tensions}} (\bibinfo {year} {2023}),\ \Eprint {https://arxiv.org/abs/2210.14339} {arXiv:2210.14339 [astro-ph.CO]} \BibitemShut {NoStop}%
\bibitem [{\citenamefont {Zhou}\ \emph {et~al.}(2022{\natexlab{b}})\citenamefont {Zhou}, \citenamefont {Liu}, \citenamefont {Mu},\ and\ \citenamefont {Xu}}]{PRD.105.103509}%
  \BibitemOpen
  \bibfield  {author} {\bibinfo {author} {\bibfnamefont {Z.}~\bibnamefont {Zhou}}, \bibinfo {author} {\bibfnamefont {G.}~\bibnamefont {Liu}}, \bibinfo {author} {\bibfnamefont {Y.}~\bibnamefont {Mu}},\ and\ \bibinfo {author} {\bibfnamefont {L.}~\bibnamefont {Xu}},\ }\bibfield  {title} {\bibinfo {title} {Limit on the dark matter mass from its interaction with photons},\ }\href {https://doi.org/10.1103/PhysRevD.105.103509} {\bibfield  {journal} {\bibinfo  {journal} {Phys. Rev. D}\ }\textbf {\bibinfo {volume} {105}},\ \bibinfo {pages} {103509} (\bibinfo {year} {2022}{\natexlab{b}})}\BibitemShut {NoStop}%
\bibitem [{\citenamefont {D'Eramo}\ \emph {et~al.}(2018)\citenamefont {D'Eramo}, \citenamefont {Ferreira}, \citenamefont {Notari},\ and\ \citenamefont {Bernal}}]{D'Eramo_2018}%
  \BibitemOpen
  \bibfield  {author} {\bibinfo {author} {\bibfnamefont {F.}~\bibnamefont {D'Eramo}}, \bibinfo {author} {\bibfnamefont {R.~Z.}\ \bibnamefont {Ferreira}}, \bibinfo {author} {\bibfnamefont {A.}~\bibnamefont {Notari}},\ and\ \bibinfo {author} {\bibfnamefont {J.~L.}\ \bibnamefont {Bernal}},\ }\bibfield  {title} {\bibinfo {title} {Hot axions and the h0 tension},\ }\href {https://doi.org/10.1088/1475-7516/2018/11/014} {\bibfield  {journal} {\bibinfo  {journal} {Journal of Cosmology and Astroparticle Physics}\ }\textbf {\bibinfo {volume} {2018}}\bibinfo  {number} { (11)},\ \bibinfo {pages} {014}}\BibitemShut {NoStop}%
\bibitem [{\citenamefont {El-Zant}\ \emph {et~al.}(2019)\citenamefont {El-Zant}, \citenamefont {Hanafy},\ and\ \citenamefont {Elgammal}}]{El-Zant_2019}%
  \BibitemOpen
\bibfield  {number} {  }\bibfield  {author} {\bibinfo {author} {\bibfnamefont {A.}~\bibnamefont {El-Zant}}, \bibinfo {author} {\bibfnamefont {W.~E.}\ \bibnamefont {Hanafy}},\ and\ \bibinfo {author} {\bibfnamefont {S.}~\bibnamefont {Elgammal}},\ }\bibfield  {title} {\bibinfo {title} {H0 tension and the phantom regime: A case study in terms of an infrared f(t) gravity},\ }\href {https://doi.org/10.3847/1538-4357/aafa12} {\bibfield  {journal} {\bibinfo  {journal} {The Astrophysical Journal}\ }\textbf {\bibinfo {volume} {871}},\ \bibinfo {pages} {210} (\bibinfo {year} {2019})}\BibitemShut {NoStop}%
\bibitem [{\citenamefont {Khosravi}\ \emph {et~al.}(2019)\citenamefont {Khosravi}, \citenamefont {Baghram}, \citenamefont {Afshordi},\ and\ \citenamefont {Altamirano}}]{PRD.99.103526}%
  \BibitemOpen
  \bibfield  {author} {\bibinfo {author} {\bibfnamefont {N.}~\bibnamefont {Khosravi}}, \bibinfo {author} {\bibfnamefont {S.}~\bibnamefont {Baghram}}, \bibinfo {author} {\bibfnamefont {N.}~\bibnamefont {Afshordi}},\ and\ \bibinfo {author} {\bibfnamefont {N.}~\bibnamefont {Altamirano}},\ }\bibfield  {title} {\bibinfo {title} {${H}_{0}$ tension as a hint for a transition in gravitational theory},\ }\href {https://doi.org/10.1103/PhysRevD.99.103526} {\bibfield  {journal} {\bibinfo  {journal} {Phys. Rev. D}\ }\textbf {\bibinfo {volume} {99}},\ \bibinfo {pages} {103526} (\bibinfo {year} {2019})}\BibitemShut {NoStop}%
\bibitem [{\citenamefont {Renk}\ \emph {et~al.}(2017)\citenamefont {Renk}, \citenamefont {Zumalacárregui}, \citenamefont {Montanari},\ and\ \citenamefont {Barreira}}]{Renk_2017}%
  \BibitemOpen
  \bibfield  {author} {\bibinfo {author} {\bibfnamefont {J.}~\bibnamefont {Renk}}, \bibinfo {author} {\bibfnamefont {M.}~\bibnamefont {Zumalacárregui}}, \bibinfo {author} {\bibfnamefont {F.}~\bibnamefont {Montanari}},\ and\ \bibinfo {author} {\bibfnamefont {A.}~\bibnamefont {Barreira}},\ }\bibfield  {title} {\bibinfo {title} {Galileon gravity in light of isw, cmb, bao and h0 data},\ }\href {https://doi.org/10.1088/1475-7516/2017/10/020} {\bibfield  {journal} {\bibinfo  {journal} {Journal of Cosmology and Astroparticle Physics}\ }\textbf {\bibinfo {volume} {2017}}\bibinfo  {number} { (10)},\ \bibinfo {pages} {020}}\BibitemShut {NoStop}%
\bibitem [{\citenamefont {Braglia}\ \emph {et~al.}(2020)\citenamefont {Braglia}, \citenamefont {Ballardini}, \citenamefont {Emond}, \citenamefont {Finelli}, \citenamefont {G\"umr\"uk\ifmmode \mbox{\c{c}}\else \c{c}\fi{}\"uo\ifmmode~\breve{g}\else \u{g}\fi{}lu}, \citenamefont {Koyama},\ and\ \citenamefont {Paoletti}}]{PRD.102.023529}%
  \BibitemOpen
\bibfield  {number} {  }\bibfield  {author} {\bibinfo {author} {\bibfnamefont {M.}~\bibnamefont {Braglia}}, \bibinfo {author} {\bibfnamefont {M.}~\bibnamefont {Ballardini}}, \bibinfo {author} {\bibfnamefont {W.~T.}\ \bibnamefont {Emond}}, \bibinfo {author} {\bibfnamefont {F.}~\bibnamefont {Finelli}}, \bibinfo {author} {\bibfnamefont {A.~E.}\ \bibnamefont {G\"umr\"uk\ifmmode \mbox{\c{c}}\else \c{c}\fi{}\"uo\ifmmode~\breve{g}\else \u{g}\fi{}lu}}, \bibinfo {author} {\bibfnamefont {K.}~\bibnamefont {Koyama}},\ and\ \bibinfo {author} {\bibfnamefont {D.}~\bibnamefont {Paoletti}},\ }\bibfield  {title} {\bibinfo {title} {Larger value for ${H}_{0}$ by an evolving gravitational constant},\ }\href {https://doi.org/10.1103/PhysRevD.102.023529} {\bibfield  {journal} {\bibinfo  {journal} {Phys. Rev. D}\ }\textbf {\bibinfo {volume} {102}},\ \bibinfo {pages} {023529} (\bibinfo {year} {2020})}\BibitemShut {NoStop}%
\bibitem [{\citenamefont {T\'ellez-Tovar}\ \emph {et~al.}(2022)\citenamefont {T\'ellez-Tovar}, \citenamefont {Matos},\ and\ \citenamefont {V\'azquez}}]{PRD.106.123501}%
  \BibitemOpen
  \bibfield  {author} {\bibinfo {author} {\bibfnamefont {L.~O.}\ \bibnamefont {T\'ellez-Tovar}}, \bibinfo {author} {\bibfnamefont {T.}~\bibnamefont {Matos}},\ and\ \bibinfo {author} {\bibfnamefont {J.~A.}\ \bibnamefont {V\'azquez}},\ }\bibfield  {title} {\bibinfo {title} {Cosmological constraints on the multiscalar field dark matter model},\ }\href {https://doi.org/10.1103/PhysRevD.106.123501} {\bibfield  {journal} {\bibinfo  {journal} {Phys. Rev. D}\ }\textbf {\bibinfo {volume} {106}},\ \bibinfo {pages} {123501} (\bibinfo {year} {2022})}\BibitemShut {NoStop}%
\bibitem [{\citenamefont {Ferreira}(2021)}]{2005.03254}%
  \BibitemOpen
  \bibfield  {author} {\bibinfo {author} {\bibfnamefont {E.~G.~M.}\ \bibnamefont {Ferreira}},\ }\bibfield  {title} {\bibinfo {title} {Ultra-light dark matter},\ }\href {https://doi.org/10.1007%2Fs00159-021-00135-6} {\bibfield  {journal} {\bibinfo  {journal} {The Astronomy and Astrophysics Review}\ }\textbf {\bibinfo {volume} {29}},\ \bibinfo {pages} {1} (\bibinfo {year} {2021})}\BibitemShut {NoStop}%
\bibitem [{\citenamefont {Ure{\~{n}}a-L{\'{o}}pez}\ and\ \citenamefont {Gonzalez-Morales}(2016)}]{Ure_a_L_pez_2016}%
  \BibitemOpen
  \bibfield  {author} {\bibinfo {author} {\bibfnamefont {L.~A.}\ \bibnamefont {Ure{\~{n}}a-L{\'{o}}pez}}\ and\ \bibinfo {author} {\bibfnamefont {A.~X.}\ \bibnamefont {Gonzalez-Morales}},\ }\bibfield  {title} {\bibinfo {title} {Towards accurate cosmological predictions for rapidly oscillating scalar fields as dark matter},\ }\href {https://doi.org/10.1088/1475-7516/2016/07/048} {\bibfield  {journal} {\bibinfo  {journal} {Journal of Cosmology and Astroparticle Physics}\ }\textbf {\bibinfo {volume} {2016}}\bibinfo  {number} { (07)},\ \bibinfo {pages} {048}}\BibitemShut {NoStop}%
\bibitem [{\citenamefont {Cede{\~{n}}o}\ \emph {et~al.}(2017)\citenamefont {Cede{\~{n}}o}, \citenamefont {Gonz{\'{a}}lez-Morales},\ and\ \citenamefont {Ure{\~{n}}a-L{\'{o}}pez}}]{PRD.96.061301}%
  \BibitemOpen
\bibfield  {number} {  }\bibfield  {author} {\bibinfo {author} {\bibfnamefont {F.~X.~L.}\ \bibnamefont {Cede{\~{n}}o}}, \bibinfo {author} {\bibfnamefont {A.~X.}\ \bibnamefont {Gonz{\'{a}}lez-Morales}},\ and\ \bibinfo {author} {\bibfnamefont {L.~A.}\ \bibnamefont {Ure{\~{n}}a-L{\'{o}}pez}},\ }\bibfield  {title} {\bibinfo {title} {Cosmological signatures of ultralight dark matter with an axionlike potential},\ }\href {https://doi.org/10.1103/PhysRevD.96.061301} {\bibfield  {journal} {\bibinfo  {journal} {Phys. Rev. D}\ }\textbf {\bibinfo {volume} {96}},\ \bibinfo {pages} {061301} (\bibinfo {year} {2017})}\BibitemShut {NoStop}%
\bibitem [{\citenamefont {Berera}(1995)}]{PRL.75.3218}%
  \BibitemOpen
  \bibfield  {author} {\bibinfo {author} {\bibfnamefont {A.}~\bibnamefont {Berera}},\ }\bibfield  {title} {\bibinfo {title} {Warm inflation},\ }\href {https://doi.org/10.1103/PhysRevLett.75.3218} {\bibfield  {journal} {\bibinfo  {journal} {Phys. Rev. Lett.}\ }\textbf {\bibinfo {volume} {75}},\ \bibinfo {pages} {3218} (\bibinfo {year} {1995})}\BibitemShut {NoStop}%
\bibitem [{\citenamefont {Berera}\ and\ \citenamefont {Fang}(1995)}]{PRL.74.1912}%
  \BibitemOpen
  \bibfield  {author} {\bibinfo {author} {\bibfnamefont {A.}~\bibnamefont {Berera}}\ and\ \bibinfo {author} {\bibfnamefont {L.-Z.}\ \bibnamefont {Fang}},\ }\bibfield  {title} {\bibinfo {title} {Thermally induced density perturbations in the inflation era},\ }\href {https://doi.org/10.1103/PhysRevLett.74.1912} {\bibfield  {journal} {\bibinfo  {journal} {Phys. Rev. Lett.}\ }\textbf {\bibinfo {volume} {74}},\ \bibinfo {pages} {1912} (\bibinfo {year} {1995})}\BibitemShut {NoStop}%
\bibitem [{\citenamefont {Berera}\ \emph {et~al.}(2009)\citenamefont {Berera}, \citenamefont {Moss},\ and\ \citenamefont {Ramos}}]{Berera_2009}%
  \BibitemOpen
  \bibfield  {author} {\bibinfo {author} {\bibfnamefont {A.}~\bibnamefont {Berera}}, \bibinfo {author} {\bibfnamefont {I.~G.}\ \bibnamefont {Moss}},\ and\ \bibinfo {author} {\bibfnamefont {R.~O.}\ \bibnamefont {Ramos}},\ }\bibfield  {title} {\bibinfo {title} {Warm inflation and its microphysical basis},\ }\href {https://doi.org/10.1088/0034-4885/72/2/026901} {\bibfield  {journal} {\bibinfo  {journal} {Reports on Progress in Physics}\ }\textbf {\bibinfo {volume} {72}},\ \bibinfo {pages} {026901} (\bibinfo {year} {2009})}\BibitemShut {NoStop}%
\bibitem [{\citenamefont {Bastero-Gil}\ \emph {et~al.}(2016)\citenamefont {Bastero-Gil}, \citenamefont {Berera}, \citenamefont {Ramos},\ and\ \citenamefont {Rosa}}]{PRL.117.151301}%
  \BibitemOpen
  \bibfield  {author} {\bibinfo {author} {\bibfnamefont {M.}~\bibnamefont {Bastero-Gil}}, \bibinfo {author} {\bibfnamefont {A.}~\bibnamefont {Berera}}, \bibinfo {author} {\bibfnamefont {R.~O.}\ \bibnamefont {Ramos}},\ and\ \bibinfo {author} {\bibfnamefont {J.~a.~G.}\ \bibnamefont {Rosa}},\ }\bibfield  {title} {\bibinfo {title} {Warm little inflaton},\ }\href {https://doi.org/10.1103/PhysRevLett.117.151301} {\bibfield  {journal} {\bibinfo  {journal} {Phys. Rev. Lett.}\ }\textbf {\bibinfo {volume} {117}},\ \bibinfo {pages} {151301} (\bibinfo {year} {2016})}\BibitemShut {NoStop}%
\bibitem [{\citenamefont {Berghaus}\ \emph {et~al.}(2020)\citenamefont {Berghaus}, \citenamefont {Graham},\ and\ \citenamefont {Kaplan}}]{Berghaus_2020}%
  \BibitemOpen
  \bibfield  {author} {\bibinfo {author} {\bibfnamefont {K.~V.}\ \bibnamefont {Berghaus}}, \bibinfo {author} {\bibfnamefont {P.~W.}\ \bibnamefont {Graham}},\ and\ \bibinfo {author} {\bibfnamefont {D.~E.}\ \bibnamefont {Kaplan}},\ }\bibfield  {title} {\bibinfo {title} {Minimal warm inflation},\ }\href {https://doi.org/10.1088/1475-7516/2020/03/034} {\bibfield  {journal} {\bibinfo  {journal} {Journal of Cosmology and Astroparticle Physics}\ }\textbf {\bibinfo {volume} {2020}}\bibinfo  {number} { (03)},\ \bibinfo {pages} {034}}\BibitemShut {NoStop}%
\bibitem [{\citenamefont {Graham}\ \emph {et~al.}(2019)\citenamefont {Graham}, \citenamefont {Kaplan},\ and\ \citenamefont {Rajendran}}]{PRD.100.015048}%
  \BibitemOpen
\bibfield  {number} {  }\bibfield  {author} {\bibinfo {author} {\bibfnamefont {P.~W.}\ \bibnamefont {Graham}}, \bibinfo {author} {\bibfnamefont {D.~E.}\ \bibnamefont {Kaplan}},\ and\ \bibinfo {author} {\bibfnamefont {S.}~\bibnamefont {Rajendran}},\ }\bibfield  {title} {\bibinfo {title} {Relaxation of the cosmological constant},\ }\href {https://doi.org/10.1103/PhysRevD.100.015048} {\bibfield  {journal} {\bibinfo  {journal} {Phys. Rev. D}\ }\textbf {\bibinfo {volume} {100}},\ \bibinfo {pages} {015048} (\bibinfo {year} {2019})}\BibitemShut {NoStop}%
\bibitem [{\citenamefont {Berghaus}\ \emph {et~al.}(2021)\citenamefont {Berghaus}, \citenamefont {Graham}, \citenamefont {Kaplan}, \citenamefont {Moore},\ and\ \citenamefont {Rajendran}}]{PRD.104.083520}%
  \BibitemOpen
  \bibfield  {author} {\bibinfo {author} {\bibfnamefont {K.~V.}\ \bibnamefont {Berghaus}}, \bibinfo {author} {\bibfnamefont {P.~W.}\ \bibnamefont {Graham}}, \bibinfo {author} {\bibfnamefont {D.~E.}\ \bibnamefont {Kaplan}}, \bibinfo {author} {\bibfnamefont {G.~D.}\ \bibnamefont {Moore}},\ and\ \bibinfo {author} {\bibfnamefont {S.}~\bibnamefont {Rajendran}},\ }\bibfield  {title} {\bibinfo {title} {Dark energy radiation},\ }\href {https://doi.org/10.1103/PhysRevD.104.083520} {\bibfield  {journal} {\bibinfo  {journal} {Phys. Rev. D}\ }\textbf {\bibinfo {volume} {104}},\ \bibinfo {pages} {083520} (\bibinfo {year} {2021})}\BibitemShut {NoStop}%
\bibitem [{\citenamefont {Berghaus}\ and\ \citenamefont {Karwal}(2020)}]{PRD.101.083537}%
  \BibitemOpen
  \bibfield  {author} {\bibinfo {author} {\bibfnamefont {K.~V.}\ \bibnamefont {Berghaus}}\ and\ \bibinfo {author} {\bibfnamefont {T.}~\bibnamefont {Karwal}},\ }\bibfield  {title} {\bibinfo {title} {Thermal friction as a solution to the hubble tension},\ }\href {https://doi.org/10.1103/PhysRevD.101.083537} {\bibfield  {journal} {\bibinfo  {journal} {Phys. Rev. D}\ }\textbf {\bibinfo {volume} {101}},\ \bibinfo {pages} {083537} (\bibinfo {year} {2020})}\BibitemShut {NoStop}%
\bibitem [{\citenamefont {Berghaus}\ and\ \citenamefont {Karwal}(2022)}]{berghaus2022thermal}%
  \BibitemOpen
  \bibfield  {author} {\bibinfo {author} {\bibfnamefont {K.~V.}\ \bibnamefont {Berghaus}}\ and\ \bibinfo {author} {\bibfnamefont {T.}~\bibnamefont {Karwal}},\ }\href@noop {} {\bibinfo {title} {Thermal friction as a solution to the hubble and large-scale structure tensions}} (\bibinfo {year} {2022}),\ \Eprint {https://arxiv.org/abs/2204.09133} {arXiv:2204.09133 [astro-ph.CO]} \BibitemShut {NoStop}%
\bibitem [{\citenamefont {Beyer}\ and\ \citenamefont {Wetterich}(2014)}]{BEYER2014418}%
  \BibitemOpen
  \bibfield  {author} {\bibinfo {author} {\bibfnamefont {J.}~\bibnamefont {Beyer}}\ and\ \bibinfo {author} {\bibfnamefont {C.}~\bibnamefont {Wetterich}},\ }\bibfield  {title} {\bibinfo {title} {Small scale structures in coupled scalar field dark matter},\ }\href {https://doi.org/https://doi.org/10.1016/j.physletb.2014.10.012} {\bibfield  {journal} {\bibinfo  {journal} {Physics Letters B}\ }\textbf {\bibinfo {volume} {738}},\ \bibinfo {pages} {418} (\bibinfo {year} {2014})}\BibitemShut {NoStop}%
\bibitem [{\citenamefont {Amendola}\ and\ \citenamefont {Barbieri}(2006)}]{AMENDOLA2006192}%
  \BibitemOpen
  \bibfield  {author} {\bibinfo {author} {\bibfnamefont {L.}~\bibnamefont {Amendola}}\ and\ \bibinfo {author} {\bibfnamefont {R.}~\bibnamefont {Barbieri}},\ }\bibfield  {title} {\bibinfo {title} {Dark matter from an ultra-light pseudo-goldsone-boson},\ }\href {https://doi.org/https://doi.org/10.1016/j.physletb.2006.08.069} {\bibfield  {journal} {\bibinfo  {journal} {Physics Letters B}\ }\textbf {\bibinfo {volume} {642}},\ \bibinfo {pages} {192} (\bibinfo {year} {2006})}\BibitemShut {NoStop}%
\bibitem [{\citenamefont {Bastero-Gil}\ \emph {et~al.}(2014)\citenamefont {Bastero-Gil}, \citenamefont {Berera}, \citenamefont {Moss},\ and\ \citenamefont {Ramos}}]{Bastero-Gil_2014}%
  \BibitemOpen
  \bibfield  {author} {\bibinfo {author} {\bibfnamefont {M.}~\bibnamefont {Bastero-Gil}}, \bibinfo {author} {\bibfnamefont {A.}~\bibnamefont {Berera}}, \bibinfo {author} {\bibfnamefont {I.~G.}\ \bibnamefont {Moss}},\ and\ \bibinfo {author} {\bibfnamefont {R.~O.}\ \bibnamefont {Ramos}},\ }\bibfield  {title} {\bibinfo {title} {Theory of non-gaussianity in warm inflation},\ }\href {https://doi.org/10.1088/1475-7516/2014/12/008} {\bibfield  {journal} {\bibinfo  {journal} {Journal of Cosmology and Astroparticle Physics}\ }\textbf {\bibinfo {volume} {2014}}\bibinfo  {number} { (12)},\ \bibinfo {pages} {008}}\BibitemShut {NoStop}%
\bibitem [{\citenamefont {McLerran}\ \emph {et~al.}(1991)\citenamefont {McLerran}, \citenamefont {Mottola},\ and\ \citenamefont {Shaposhnikov}}]{PRD.43.2027}%
  \BibitemOpen
\bibfield  {number} {  }\bibfield  {author} {\bibinfo {author} {\bibfnamefont {L.}~\bibnamefont {McLerran}}, \bibinfo {author} {\bibfnamefont {E.}~\bibnamefont {Mottola}},\ and\ \bibinfo {author} {\bibfnamefont {M.~E.}\ \bibnamefont {Shaposhnikov}},\ }\bibfield  {title} {\bibinfo {title} {Sphalerons and axion dynamics in high-temperature qcd},\ }\href {https://doi.org/10.1103/PhysRevD.43.2027} {\bibfield  {journal} {\bibinfo  {journal} {Phys. Rev. D}\ }\textbf {\bibinfo {volume} {43}},\ \bibinfo {pages} {2027} (\bibinfo {year} {1991})}\BibitemShut {NoStop}%
\bibitem [{\citenamefont {Moore}\ and\ \citenamefont {Tassler}(2011)}]{Moore_2011}%
  \BibitemOpen
  \bibfield  {author} {\bibinfo {author} {\bibfnamefont {G.~D.}\ \bibnamefont {Moore}}\ and\ \bibinfo {author} {\bibfnamefont {M.}~\bibnamefont {Tassler}},\ }\bibfield  {title} {\bibinfo {title} {The sphaleron rate in {SU}(n) gauge theory},\ }\bibfield  {journal} {\bibinfo  {journal} {Journal of High Energy Physics}\ }\textbf {\bibinfo {volume} {2011}},\ \href {https://doi.org/10.1007/jhep02(2011)105} {10.1007/jhep02(2011)105} (\bibinfo {year} {2011})\BibitemShut {NoStop}%
\bibitem [{\citenamefont {Laine}\ and\ \citenamefont {Vuorinen}(2016)}]{Laine_2016}%
  \BibitemOpen
  \bibfield  {author} {\bibinfo {author} {\bibfnamefont {M.}~\bibnamefont {Laine}}\ and\ \bibinfo {author} {\bibfnamefont {A.}~\bibnamefont {Vuorinen}},\ }\href {https://doi.org/10.1007/978-3-319-31933-9} {\emph {\bibinfo {title} {Basics of Thermal Field Theory}}}\ (\bibinfo  {publisher} {Springer International Publishing},\ \bibinfo {year} {2016})\BibitemShut {NoStop}%
\bibitem [{\citenamefont {Copeland}\ \emph {et~al.}(1998)\citenamefont {Copeland}, \citenamefont {Liddle},\ and\ \citenamefont {Wands}}]{PRD.57.4686}%
  \BibitemOpen
  \bibfield  {author} {\bibinfo {author} {\bibfnamefont {E.~J.}\ \bibnamefont {Copeland}}, \bibinfo {author} {\bibfnamefont {A.~R.}\ \bibnamefont {Liddle}},\ and\ \bibinfo {author} {\bibfnamefont {D.}~\bibnamefont {Wands}},\ }\bibfield  {title} {\bibinfo {title} {Exponential potentials and cosmological scaling solutions},\ }\href {https://doi.org/10.1103/PhysRevD.57.4686} {\bibfield  {journal} {\bibinfo  {journal} {Phys. Rev. D}\ }\textbf {\bibinfo {volume} {57}},\ \bibinfo {pages} {4686} (\bibinfo {year} {1998})}\BibitemShut {NoStop}%
\bibitem [{\citenamefont {Audren}\ \emph {et~al.}(2014)\citenamefont {Audren}, \citenamefont {Lesgourgues}, \citenamefont {Mangano}, \citenamefont {Serpico},\ and\ \citenamefont {Tram}}]{Audren_2014}%
  \BibitemOpen
  \bibfield  {author} {\bibinfo {author} {\bibfnamefont {B.}~\bibnamefont {Audren}}, \bibinfo {author} {\bibfnamefont {J.}~\bibnamefont {Lesgourgues}}, \bibinfo {author} {\bibfnamefont {G.}~\bibnamefont {Mangano}}, \bibinfo {author} {\bibfnamefont {P.~D.}\ \bibnamefont {Serpico}},\ and\ \bibinfo {author} {\bibfnamefont {T.}~\bibnamefont {Tram}},\ }\bibfield  {title} {\bibinfo {title} {Strongest model-independent bound on the lifetime of dark matter},\ }\href {https://doi.org/10.1088/1475-7516/2014/12/028} {\bibfield  {journal} {\bibinfo  {journal} {Journal of Cosmology and Astroparticle Physics}\ }\textbf {\bibinfo {volume} {2014}}\bibinfo  {number} { (12)},\ \bibinfo {pages} {028}}\BibitemShut {NoStop}%
\bibitem [{\citenamefont {Lesgourgues}\ \emph {et~al.}(2016)\citenamefont {Lesgourgues}, \citenamefont {Marques-Tavares},\ and\ \citenamefont {Schmaltz}}]{Lesgourgues_2016}%
  \BibitemOpen
\bibfield  {number} {  }\bibfield  {author} {\bibinfo {author} {\bibfnamefont {J.}~\bibnamefont {Lesgourgues}}, \bibinfo {author} {\bibfnamefont {G.}~\bibnamefont {Marques-Tavares}},\ and\ \bibinfo {author} {\bibfnamefont {M.}~\bibnamefont {Schmaltz}},\ }\bibfield  {title} {\bibinfo {title} {Evidence for dark matter interactions in cosmological precision data?},\ }\href {https://doi.org/10.1088/1475-7516/2016/02/037} {\bibfield  {journal} {\bibinfo  {journal} {Journal of Cosmology and Astroparticle Physics}\ }\textbf {\bibinfo {volume} {2016}}\bibinfo  {number} { (02)},\ \bibinfo {pages} {037}}\BibitemShut {NoStop}%
\bibitem [{\citenamefont {Ferreira}\ and\ \citenamefont {Joyce}(1998)}]{PRD.58.023503}%
  \BibitemOpen
\bibfield  {number} {  }\bibfield  {author} {\bibinfo {author} {\bibfnamefont {P.~G.}\ \bibnamefont {Ferreira}}\ and\ \bibinfo {author} {\bibfnamefont {M.}~\bibnamefont {Joyce}},\ }\bibfield  {title} {\bibinfo {title} {Cosmology with a primordial scaling field},\ }\href {https://doi.org/10.1103/PhysRevD.58.023503} {\bibfield  {journal} {\bibinfo  {journal} {Phys. Rev. D}\ }\textbf {\bibinfo {volume} {58}},\ \bibinfo {pages} {023503} (\bibinfo {year} {1998})}\BibitemShut {NoStop}%
\bibitem [{\citenamefont {Hu}(1998)}]{Hu_1998}%
  \BibitemOpen
  \bibfield  {author} {\bibinfo {author} {\bibfnamefont {W.}~\bibnamefont {Hu}},\ }\bibfield  {title} {\bibinfo {title} {Structure formation with generalized dark matter},\ }\href {https://doi.org/10.1086/306274} {\bibfield  {journal} {\bibinfo  {journal} {The Astrophysical Journal}\ }\textbf {\bibinfo {volume} {506}},\ \bibinfo {pages} {485} (\bibinfo {year} {1998})}\BibitemShut {NoStop}%
\bibitem [{\citenamefont {Poulin}\ \emph {et~al.}(2016)\citenamefont {Poulin}, \citenamefont {Serpico},\ and\ \citenamefont {Lesgourgues}}]{Poulin_2016}%
  \BibitemOpen
  \bibfield  {author} {\bibinfo {author} {\bibfnamefont {V.}~\bibnamefont {Poulin}}, \bibinfo {author} {\bibfnamefont {P.~D.}\ \bibnamefont {Serpico}},\ and\ \bibinfo {author} {\bibfnamefont {J.}~\bibnamefont {Lesgourgues}},\ }\bibfield  {title} {\bibinfo {title} {A fresh look at linear cosmological constraints on a decaying dark matter component},\ }\href {https://doi.org/10.1088/1475-7516/2016/08/036} {\bibfield  {journal} {\bibinfo  {journal} {Journal of Cosmology and Astroparticle Physics}\ }\textbf {\bibinfo {volume} {2016}}\bibinfo  {number} { (08)},\ \bibinfo {pages} {036}}\BibitemShut {NoStop}%
\bibitem [{\citenamefont {Simon}\ \emph {et~al.}(2022)\citenamefont {Simon}, \citenamefont {Abell\'an}, \citenamefont {Du}, \citenamefont {Poulin},\ and\ \citenamefont {Tsai}}]{PRD.106.023516}%
  \BibitemOpen
\bibfield  {number} {  }\bibfield  {author} {\bibinfo {author} {\bibfnamefont {T.}~\bibnamefont {Simon}}, \bibinfo {author} {\bibfnamefont {G.~F.}\ \bibnamefont {Abell\'an}}, \bibinfo {author} {\bibfnamefont {P.}~\bibnamefont {Du}}, \bibinfo {author} {\bibfnamefont {V.}~\bibnamefont {Poulin}},\ and\ \bibinfo {author} {\bibfnamefont {Y.}~\bibnamefont {Tsai}},\ }\bibfield  {title} {\bibinfo {title} {Constraining decaying dark matter with boss data and the effective field theory of large-scale structures},\ }\href {https://doi.org/10.1103/PhysRevD.106.023516} {\bibfield  {journal} {\bibinfo  {journal} {Phys. Rev. D}\ }\textbf {\bibinfo {volume} {106}},\ \bibinfo {pages} {023516} (\bibinfo {year} {2022})}\BibitemShut {NoStop}%
\bibitem [{\citenamefont {Lesgourgues}(2011)}]{1104.2932}%
  \BibitemOpen
  \bibfield  {author} {\bibinfo {author} {\bibfnamefont {J.}~\bibnamefont {Lesgourgues}},\ }\bibfield  {title} {\bibinfo {title} {The cosmic linear anisotropy solving system (class) i: Overview}\ }(\bibinfo {year} {2011})\ \Eprint {https://arxiv.org/abs/1104.2932} {arXiv:1104.2932 [astro-ph.IM]} \BibitemShut {NoStop}%
\bibitem [{\citenamefont {Blas}\ \emph {et~al.}(2011)\citenamefont {Blas}, \citenamefont {Lesgourgues},\ and\ \citenamefont {Tram}}]{Blas_2011}%
  \BibitemOpen
  \bibfield  {author} {\bibinfo {author} {\bibfnamefont {D.}~\bibnamefont {Blas}}, \bibinfo {author} {\bibfnamefont {J.}~\bibnamefont {Lesgourgues}},\ and\ \bibinfo {author} {\bibfnamefont {T.}~\bibnamefont {Tram}},\ }\bibfield  {title} {\bibinfo {title} {The cosmic linear anisotropy solving system ({CLASS}). part {II}: Approximation schemes},\ }\href {https://doi.org/10.1088/1475-7516/2011/07/034} {\bibfield  {journal} {\bibinfo  {journal} {Journal of Cosmology and Astroparticle Physics}\ }\textbf {\bibinfo {volume} {2011}}\bibinfo  {number} { (07)},\ \bibinfo {pages} {034}}\BibitemShut {NoStop}%
\bibitem [{\citenamefont {Torrado}\ and\ \citenamefont {Lewis}(2021)}]{Torrado_2021}%
  \BibitemOpen
\bibfield  {number} {  }\bibfield  {author} {\bibinfo {author} {\bibfnamefont {J.}~\bibnamefont {Torrado}}\ and\ \bibinfo {author} {\bibfnamefont {A.}~\bibnamefont {Lewis}},\ }\bibfield  {title} {\bibinfo {title} {Cobaya: code for bayesian analysis of hierarchical physical models},\ }\href {https://doi.org/10.1088/1475-7516/2021/05/057} {\bibfield  {journal} {\bibinfo  {journal} {Journal of Cosmology and Astroparticle Physics}\ }\textbf {\bibinfo {volume} {2021}}\bibinfo  {number} { (05)},\ \bibinfo {pages} {057}}\BibitemShut {NoStop}%
\bibitem [{\citenamefont {Gelman}\ and\ \citenamefont {Rubin}(1992)}]{Gelman1992InferenceFI}%
  \BibitemOpen
\bibfield  {number} {  }\bibfield  {author} {\bibinfo {author} {\bibfnamefont {A.}~\bibnamefont {Gelman}}\ and\ \bibinfo {author} {\bibfnamefont {D.~B.}\ \bibnamefont {Rubin}},\ }\bibfield  {title} {\bibinfo {title} {Inference from iterative simulation using multiple sequences},\ }\href {http://dx.doi.org/ 10.1214/ss/1177011136} {\bibfield  {journal} {\bibinfo  {journal} {Statistical Science}\ }\textbf {\bibinfo {volume} {7}},\ \bibinfo {pages} {457} (\bibinfo {year} {1992})}\BibitemShut {NoStop}%
\bibitem [{\citenamefont {Lewis}(2019)}]{lewis2019getdist}%
  \BibitemOpen
  \bibfield  {author} {\bibinfo {author} {\bibfnamefont {A.}~\bibnamefont {Lewis}},\ }\href@noop {} {\bibinfo {title} {Getdist: a python package for analysing monte carlo samples}} (\bibinfo {year} {2019}),\ \Eprint {https://arxiv.org/abs/1910.13970} {arXiv:1910.13970 [astro-ph.IM]} \BibitemShut {NoStop}%
\bibitem [{\citenamefont {Aghanim}\ \emph {et~al.}(2020{\natexlab{a}})\citenamefont {Aghanim}, \citenamefont {Akrami}, \citenamefont {Ashdown} \emph {et~al.}}]{osti_1676388}%
  \BibitemOpen
  \bibfield  {author} {\bibinfo {author} {\bibfnamefont {N.}~\bibnamefont {Aghanim}}, \bibinfo {author} {\bibfnamefont {Y.}~\bibnamefont {Akrami}}, \bibinfo {author} {\bibfnamefont {M.}~\bibnamefont {Ashdown}}, \emph {et~al.},\ }\bibfield  {title} {\bibinfo {title} {Planck 2018 results. v. cmb power spectra and likelihoods},\ }\bibfield  {journal} {\bibinfo  {journal} {Astronomy and Astrophysics}\ }\textbf {\bibinfo {volume} {641}},\ \href {https://doi.org/10.1051/0004-6361/201936386} {10.1051/0004-6361/201936386} (\bibinfo {year} {2020}{\natexlab{a}})\BibitemShut {NoStop}%
\bibitem [{\citenamefont {Aghanim}\ \emph {et~al.}(2020{\natexlab{b}})\citenamefont {Aghanim}, \citenamefont {Akrami}, \citenamefont {Ashdown} \emph {et~al.}}]{osti_1775409}%
  \BibitemOpen
  \bibfield  {author} {\bibinfo {author} {\bibfnamefont {N.}~\bibnamefont {Aghanim}}, \bibinfo {author} {\bibfnamefont {Y.}~\bibnamefont {Akrami}}, \bibinfo {author} {\bibfnamefont {M.}~\bibnamefont {Ashdown}}, \emph {et~al.},\ }\bibfield  {title} {\bibinfo {title} {Planck 2018 results - viii. gravitational lensing},\ }\bibfield  {journal} {\bibinfo  {journal} {Astronomy and Astrophysics}\ }\textbf {\bibinfo {volume} {641}},\ \href {https://doi.org/10.1051/0004-6361/201833886} {10.1051/0004-6361/201833886} (\bibinfo {year} {2020}{\natexlab{b}})\BibitemShut {NoStop}%
\bibitem [{\citenamefont {Alam}\ \emph {et~al.}(2017)\citenamefont {Alam}, \citenamefont {Ata}, \citenamefont {Bailey} \emph {et~al.}}]{Alam_2017}%
  \BibitemOpen
  \bibfield  {author} {\bibinfo {author} {\bibfnamefont {S.}~\bibnamefont {Alam}}, \bibinfo {author} {\bibfnamefont {M.}~\bibnamefont {Ata}}, \bibinfo {author} {\bibfnamefont {S.}~\bibnamefont {Bailey}}, \emph {et~al.},\ }\bibfield  {title} {\bibinfo {title} {The clustering of galaxies in the completed {SDSS}-{III} baryon oscillation spectroscopic survey: cosmological analysis of the {DR}12 galaxy sample},\ }\href {https://doi.org/10.1093/mnras/stx721} {\bibfield  {journal} {\bibinfo  {journal} {Monthly Notices of the Royal Astronomical Society}\ }\textbf {\bibinfo {volume} {470}},\ \bibinfo {pages} {2617} (\bibinfo {year} {2017})}\BibitemShut {NoStop}%
\bibitem [{\citenamefont {Buen-Abad}\ \emph {et~al.}(2018)\citenamefont {Buen-Abad}, \citenamefont {Schmaltz}, \citenamefont {Lesgourgues},\ and\ \citenamefont {Brinckmann}}]{Buen_Abad_2018}%
  \BibitemOpen
  \bibfield  {author} {\bibinfo {author} {\bibfnamefont {M.~A.}\ \bibnamefont {Buen-Abad}}, \bibinfo {author} {\bibfnamefont {M.}~\bibnamefont {Schmaltz}}, \bibinfo {author} {\bibfnamefont {J.}~\bibnamefont {Lesgourgues}},\ and\ \bibinfo {author} {\bibfnamefont {T.}~\bibnamefont {Brinckmann}},\ }\bibfield  {title} {\bibinfo {title} {Interacting dark sector and precision cosmology},\ }\href {https://doi.org/10.1088/1475-7516/2018/01/008} {\bibfield  {journal} {\bibinfo  {journal} {Journal of Cosmology and Astroparticle Physics}\ }\textbf {\bibinfo {volume} {2018}}\bibinfo  {number} { (01)},\ \bibinfo {pages} {008}}\BibitemShut {NoStop}%
\bibitem [{\citenamefont {Beutler}\ \emph {et~al.}(2011)\citenamefont {Beutler}, \citenamefont {Blake}, \citenamefont {Colless} \emph {et~al.}}]{19250.x}%
  \BibitemOpen
\bibfield  {number} {  }\bibfield  {author} {\bibinfo {author} {\bibfnamefont {F.}~\bibnamefont {Beutler}}, \bibinfo {author} {\bibfnamefont {C.}~\bibnamefont {Blake}}, \bibinfo {author} {\bibfnamefont {M.}~\bibnamefont {Colless}}, \emph {et~al.},\ }\bibfield  {title} {\bibinfo {title} {The 6df galaxy survey: baryon acoustic oscillations and the local hubble constant},\ }\href {https://doi.org/10.1111/j.1365-2966.2011.19250.x} {\bibfield  {journal} {\bibinfo  {journal} {Monthly Notices of the Royal Astronomical Society}\ }\textbf {\bibinfo {volume} {416}},\ \bibinfo {pages} {3017} (\bibinfo {year} {2011})}\BibitemShut {NoStop}%
\bibitem [{\citenamefont {Ross}\ \emph {et~al.}(2015)\citenamefont {Ross}, \citenamefont {Samushia}, \citenamefont {Howlett} \emph {et~al.}}]{stv154}%
  \BibitemOpen
  \bibfield  {author} {\bibinfo {author} {\bibfnamefont {A.~J.}\ \bibnamefont {Ross}}, \bibinfo {author} {\bibfnamefont {L.}~\bibnamefont {Samushia}}, \bibinfo {author} {\bibfnamefont {C.}~\bibnamefont {Howlett}}, \emph {et~al.},\ }\bibfield  {title} {\bibinfo {title} {The clustering of the sdss dr7 main galaxy sample-i. a 4 percent distance measure at z=0.15},\ }\href {https://doi.org/10.1093/mnras/stv154} {\bibfield  {journal} {\bibinfo  {journal} {Monthly Notices of the Royal Astronomical Society}\ }\textbf {\bibinfo {volume} {449}},\ \bibinfo {pages} {835} (\bibinfo {year} {2015})}\BibitemShut {NoStop}%
\bibitem [{\citenamefont {Scolnic}\ \emph {et~al.}(2018)\citenamefont {Scolnic}, \citenamefont {Jones}, \citenamefont {Rest}, \citenamefont {Pan} \emph {et~al.}}]{Scolnic_2018}%
  \BibitemOpen
  \bibfield  {author} {\bibinfo {author} {\bibfnamefont {D.~M.}\ \bibnamefont {Scolnic}}, \bibinfo {author} {\bibfnamefont {D.~O.}\ \bibnamefont {Jones}}, \bibinfo {author} {\bibfnamefont {A.}~\bibnamefont {Rest}}, \bibinfo {author} {\bibfnamefont {Y.~C.}\ \bibnamefont {Pan}}, \emph {et~al.},\ }\bibfield  {title} {\bibinfo {title} {The complete light-curve sample of spectroscopically confirmed {SNe} ia from pan-{STARRS}1 and cosmological constraints from the combined pantheon sample},\ }\href {https://doi.org/10.3847/1538-4357/aab9bb} {\bibfield  {journal} {\bibinfo  {journal} {The Astrophysical Journal}\ }\textbf {\bibinfo {volume} {859}},\ \bibinfo {pages} {101} (\bibinfo {year} {2018})}\BibitemShut {NoStop}%
\bibitem [{\citenamefont {Smith}\ \emph {et~al.}(2021)\citenamefont {Smith}, \citenamefont {Poulin}, \citenamefont {Bernal}, \citenamefont {Boddy}, \citenamefont {Kamionkowski},\ and\ \citenamefont {Murgia}}]{PRD.103.123542}%
  \BibitemOpen
  \bibfield  {author} {\bibinfo {author} {\bibfnamefont {T.~L.}\ \bibnamefont {Smith}}, \bibinfo {author} {\bibfnamefont {V.}~\bibnamefont {Poulin}}, \bibinfo {author} {\bibfnamefont {J.~L.}\ \bibnamefont {Bernal}}, \bibinfo {author} {\bibfnamefont {K.~K.}\ \bibnamefont {Boddy}}, \bibinfo {author} {\bibfnamefont {M.}~\bibnamefont {Kamionkowski}},\ and\ \bibinfo {author} {\bibfnamefont {R.}~\bibnamefont {Murgia}},\ }\bibfield  {title} {\bibinfo {title} {Early dark energy is not excluded by current large-scale structure data},\ }\href {https://doi.org/10.1103/PhysRevD.103.123542} {\bibfield  {journal} {\bibinfo  {journal} {Phys. Rev. D}\ }\textbf {\bibinfo {volume} {103}},\ \bibinfo {pages} {123542} (\bibinfo {year} {2021})}\BibitemShut {NoStop}%
\bibitem [{\citenamefont {Chabanier}\ \emph {et~al.}(2019)\citenamefont {Chabanier}, \citenamefont {Palanque-Delabrouille}, \citenamefont {Yèche} \emph {et~al.}}]{Chabanier_2019}%
  \BibitemOpen
  \bibfield  {author} {\bibinfo {author} {\bibfnamefont {S.}~\bibnamefont {Chabanier}}, \bibinfo {author} {\bibfnamefont {N.}~\bibnamefont {Palanque-Delabrouille}}, \bibinfo {author} {\bibfnamefont {C.}~\bibnamefont {Yèche}}, \emph {et~al.},\ }\bibfield  {title} {\bibinfo {title} {The one-dimensional power spectrum from the sdss dr14 ly$\alpha$ forests},\ }\href {https://doi.org/10.1088/1475-7516/2019/07/017} {\bibfield  {journal} {\bibinfo  {journal} {Journal of Cosmology and Astroparticle Physics}\ }\textbf {\bibinfo {volume} {2019}}\bibinfo  {number} { (07)},\ \bibinfo {pages} {017}}\BibitemShut {NoStop}%
\bibitem [{\citenamefont {Goldstein}\ \emph {et~al.}(2023)\citenamefont {Goldstein}, \citenamefont {Hill}, \citenamefont {Ir\ifmmode \check{s}\else \v{s}\fi{}i\ifmmode~\check{c}\else \v{c}\fi{}},\ and\ \citenamefont {Sherwin}}]{goldstein2023canonical}%
  \BibitemOpen
\bibfield  {number} {  }\bibfield  {author} {\bibinfo {author} {\bibfnamefont {S.}~\bibnamefont {Goldstein}}, \bibinfo {author} {\bibfnamefont {J.~C.}\ \bibnamefont {Hill}}, \bibinfo {author} {\bibfnamefont {V.}~\bibnamefont {Ir\ifmmode \check{s}\else \v{s}\fi{}i\ifmmode~\check{c}\else \v{c}\fi{}}},\ and\ \bibinfo {author} {\bibfnamefont {B.~D.}\ \bibnamefont {Sherwin}},\ }\bibfield  {title} {\bibinfo {title} {Canonical hubble-tension-resolving early dark energy cosmologies are inconsistent with the lyman-$\ensuremath{\alpha}$ forest},\ }\href {https://doi.org/10.1103/PhysRevLett.131.201001} {\bibfield  {journal} {\bibinfo  {journal} {Phys. Rev. Lett.}\ }\textbf {\bibinfo {volume} {131}},\ \bibinfo {pages} {201001} (\bibinfo {year} {2023})}\BibitemShut {NoStop}%
\bibitem [{\citenamefont {He}\ \emph {et~al.}(2023)\citenamefont {He}, \citenamefont {An}, \citenamefont {Ivanov},\ and\ \citenamefont {Gluscevic}}]{he2023selfinteracting}%
  \BibitemOpen
  \bibfield  {author} {\bibinfo {author} {\bibfnamefont {A.}~\bibnamefont {He}}, \bibinfo {author} {\bibfnamefont {R.}~\bibnamefont {An}}, \bibinfo {author} {\bibfnamefont {M.~M.}\ \bibnamefont {Ivanov}},\ and\ \bibinfo {author} {\bibfnamefont {V.}~\bibnamefont {Gluscevic}},\ }\href@noop {} {\bibinfo {title} {Self-interacting neutrinos in light of large-scale structure data}} (\bibinfo {year} {2023}),\ \Eprint {https://arxiv.org/abs/2309.03956} {arXiv:2309.03956 [astro-ph.CO]} \BibitemShut {NoStop}%
\end{thebibliography}%

\end{document}